\documentclass[a4paper,APS,showpacs,preprintnumbers,nofootinbib]{revtex4}
\usepackage{amssymb}
\usepackage{amsmath}
\usepackage{graphics}
\usepackage{graphicx}
\usepackage{dcolumn}
\usepackage{bm}
\usepackage{color}
\usepackage[T1]{fontenc}
\usepackage[utf8]{inputenc}
\usepackage{epstopdf}
\usepackage{ulem}
\begin{document}

\newcommand{\blue}[1]{\textcolor{blue}{#1}}
\newcommand{\new}{\blue}
\newcommand{\green}[1]{\textcolor{green}{#1}}
\newcommand{\modif}{\green}
\newcommand{\red}[1]{\textcolor{red}{#1}}
\newcommand{\attention}{\red}

\title{The Tsallis Entropy and the BKT--like Phase Transition in the Impact Parameter Space for $pp$ and $\bar{p}p$ Collisions}

\author{S. D. Campos} \email{sergiodc@ufscar.br}
\author{V. A. Okorokov} \email{VAOkorokov@mephi.ru; Okorokov@bnl.gov}
\author{C. V. Moraes$^{*}$}
\affiliation{$^{*}$Departamento de F\'isica, Qu\'imica e
Matem\'atica, Universidade Federal de S\~ao Carlos, 18052-780,
Sorocaba, SP, Brazil} \affiliation{$^{\dagger}$National Research
Nuclear University MEPhI (Moscow Engineering Physics Institute),
Kashirskoe highway 31, 115409 Moscow, Russia}

\date{\today}

\begin{abstract}
In this paper, one uses the Tsallis entropy in the impact parameter space to study $pp$ and $\bar{p}p$ inelastic overlap function and the energy density filling up mechanism responsible by the so-called black disk limit as the energy increases. The Tsallis entropy is non-additive and non-extensive and these features are of fundamental importance since the internal constituents of $pp$ and $\bar{p}p$ are strongly correlated and also the existence of the  multifractal character of the total cross-section. The entropy approach presented here takes into account a phase transition occurring inside the hadrons as the energy increases. This phase transition in the impact parameter space is quite similar to the Berezinskii--Kosterlitz--Thouless phase transition, possessing also a topological feature due to the multifractal dimension of the total cross-sections in $pp$ and $\bar{p}p$ scattering.

\end{abstract}


\maketitle

%%%%%%%%%%%%%%%%%%%%%%%%%%%%%%%%%%%%%%%%%%%%%%%%%%%%%%%%%%%%%%%%%%%%%%%%
\section{Introduction}

The transition point separating two different states of matter is known as a phase transition and is observed when a physical system suddenly changes its macroscopic behavior due to a smooth variation of the order parameter passing through its critical value. In the classical point of view, the temperature is usually the driving parameter of such transitions and determines exactly the phase transition by heat transfer. Both the superfluid helium and the Ginzburg--Landau superconducting model are very well-known examples where phase transitions are thermal-dependent in the classical sense. In the quantum world at zero temperature, the tuning parameter is chosen depending on the experiment (for instance, the magnetic field, chemical potential or electric field), and in these systems, the phase transition is known as the quantum phase transition \cite{sachdev_book_2011}.

In a modern view, phase transitions (classical or quantum) are classified as first-order or second-order. The first-order phase transition is determined by the discontinuity at first derivative of a relevant thermodynamic potential and, as consequence, such phase transition can also be called as discontinuous one, from the point of view of the behavior for the first
derivative of the thermodynamic potential. The second-order phase transition is continuous at the first derivative and divergent at higher order derivatives, and such phase transition can also be called as continuous one. Despite the significant progress in the study of critical phenomena there some disagreements with regard to terminology. For instance, the classification of phase transitions as first-order or continuous is justified in \cite{fisher_rep_prog_phys_30_615_1967}. However, such classification seems self-contradictory because uses different categories, namely, "order" and the feature (continuity/discontinuity) in the behavior of the first derivative of a relevant thermodynamic potential. Therefore the well-known classification as first- /second-order \cite{Landau-StatisticalPhysics-1980} is used for phase transitions in this paper. In any classification, the so-called critical point
separates the system ordering from a symmetric state to a broken-symmetry state.

The Berezinskii--Kosterlitz--Thouless (BKT) phase transition \cite{berezinskii_zh_eksp_teor_fiz_59_907_1970,kosterliz_j_phys_c6_1181_1973}, whose main goal is the study of correlations between pairs of topological defects (vortex-antivortex), is an example from a geometrical point of view. In condensed matter physics, the study of such transition is a tool to understand how collective coherent phenomena (a topological defect) can emerge from certain structures and how it can be connected to another one in a different space. If the interaction between topological defects depends on the logarithm of the spatial separation, then the BKT phase transition takes place being used to explain the phenomenon. As will show later, the model proposed here presents a BKT-like phase transition.

As well-known, phase transitions are closely connected to the entropy of a system and to the Helmholtz free energy, which is nothing more than the useful work that can be extracted from a closed thermodynamic system at a constant temperature. In general, one applies the Boltzmann entropy (extensive) to this kind of systems. However, a non-extensive form of entropy based on the $q$-Gaussian was proposed and successfully used to calculate and explain some physical properties in several systems, in particular, those with fractal properties. This is the so-called Tsallis entropy (TE) \cite{tsallis}. Notice that in the last decades, various approaches have been used to calculate the entropy of systems, presenting interesting properties. The Shannon entropy \cite{shannon}, the R\'enyi entropy \cite{renyi}, and the von Neumann entropy \cite{neumann} are examples of such calculation approaches, each one applied to a specific physical problem. However, all these formulations can be reduced to the TE \cite{beck_0902.1235v2}.

Entropy is one of the most important physical quantities in thermodynamics and cannot be put aside in any reliable model, even when its results are disconcerting in the classical world \cite{nature}. Moreover, entropy cannot only be viewed as the disorder of a given system. Since the fundamental work of Shannon \cite{shannon}, the entropy have been also related to the amount of information we can attain from the system. It should be stressed that the applicability of such quantity as "entropy" for the study of an interaction process between atomic nuclei and particles requires a detailed and rigorous justification. Here one can note the following with regard of two main problems: (i) a finite number of particles in the system under consideration and (ii) the time $\mathcal{T}$ invariance in the quantum field theory (QFT). First of all, the usual way to use entropy is by assuming some large statistical ensemble (canonical, for instance). In the classical view, the number of elements in such an ensemble is at least of the order of the Avogadro number. However, there are filtering methods used in information theory to reduce the number of bits, allowing at least an estimation to the entropy. Then, the size of the ensemble is not a problem, at first glance. Of course, if the size of the ensemble grows, then the estimation may also tend to a "better" value \cite{Entropy-19-520-2017}. Second, indeed, the entropy almost always grows. Nevertheless, there are some systems where the entropy achieves negative values. The concept of negative absolute temperature can be applied in such systems \cite{nature}. In these systems, the arrow of time is the same, but a non-trivial interpretation must be used to explain the result. So, the $\mathcal{T}$ invariance of the QFT is preserved and, in particular, the strong interactions are in a safe place.

The calculation of the entropy is a complicated task but can provide remarkable results as the proton and electron radii in the nucleus \cite{jimenez}. The $z$-scaling also can be used to furnish the entropy in a system with particle production \cite{zborovsky_int_j_mod_phys_a24_1417_2009} and further be calculated in $D$-branes \cite{vancea}. All these approaches used to evaluate the entropy reveal some particular thermodynamics aspect of the system under study.

Recently, the concept of fractal dimensions was introduced in the study of $pp$ and $\bar{p}p$ total cross sections \cite{borcsik_mod_phys_lett_a31_1650066_2016,okorokov_int_j_mod_phys_a32_1750175_2017}. These fractal dimensions were used to explain the transition from a decreasing total cross section for an increasing one as the energy tends to infinity. The fractal dimensions emerging in these pictures are energy-dependent in the sense that for the decreasing total cross section as the energy grows up to a critical value one obtains a negative fractal dimension, and for the increasing total cross section, one has a positive fractal dimension. On the other hand, in terms of momentum space (or in configuration space), an attempt to connect the intermittency pattern \cite{A.Bialas.R.Peschanski.Nucl.Phys.B273.703.1986,A.Bialas.R.Peschanki.Nucl.Phys.B308.857.1988,R.C.Hwa.Phys.Rev.D41.1456.1990,bialas_1,bialas_2} in hadronic collisions to fractal dimensions uses a second-order phase transition \cite{antoniou_1,antoniou_2,antoniou_4,antoniou_5}. The approach presented in \cite{deppman_phys_rev_d93_054001_2016} shows this possibility through the use of thermofractals. It should be stressed, however, that in the energy-momentum space a system with a particular fractal structure can also be described by the Tsallis statistics. In particular, this system is scale-free and can be used to study the similarities between the nonextensitiy in hadron systems and the Hagedorn's fireballs, for instance. The intermittency effects can be used to study fractal-like properties of several systems as, for instance, the multiparticle production \cite{I.Zborovsk.M.V.Tokarev.Phys.Rev.D75.094008.2007,G.Wilk.Z.Wlodarczyk.Phys.Lett.B727.163.2013}, and the gluon emission resulting from the QCD evolution equation \cite{G.Altarelli.G.Parisi.Nucl.Phys.B126.298.1977}. The fractal structure in the energy-momentum space can also be used to study scaling properties through the Callan-Symanzik renormalization group equation \cite{A.Deppman.T.Frederico.E.Megias.D.P.Menezes.Entropy.20.633.2018,A.Deppman.Adv.High.Ener.Phys.2018.9141249.2018}.

In the present paper, is proposed a naive model to evaluate the entropy of $pp$ and $\bar{p}p$ elastic scattering adopting subtleties assumptions. These assertions allow the connection of the TE and the inelastic overlap function in the impact parameter space, providing a novel interpretation of the energy density filling up mechanism of the hadron as the collision energy increases. This process can enhance the understanding of how the black disk limit is achieved (or not).

The paper is organized as follows. In section \ref{ip} the impact parameter space basic formalism is presented. In section \ref{bkt} only the essential of the TE is presented as well as our model. Section \ref{comp} present a basic example of an application using the most general experimental results and phenomenological approaches to the inelastic overlap function. Section \ref{fr} presents critical remarks about the results.

%%%%%%%%%%%%%%%%%%%%%%%%%%%%%%%%%
\section{\label{ip}Impact Parameter Point of View}

The impact parameter space is the right place to study the fractal behavior of $pp$ and $\bar{p}p$ since it allows a general view of the elastic and inelastic scattering channels. In this way, using the impact parameter formalism, the fractal dimensions obtained in \cite{borcsik_mod_phys_lett_a31_1650066_2016,antoniou_1,antoniou_2,antoniou_4,antoniou_5,bialas_1,bialas_2} may be viewed as a consequence of a phase transition in $pp$ and $\bar{p}p$ elastic scattering indicating a geometric phase transition. This topological phase transition taking place inside the hadron may be responsible be a change in the energy density filling up mechanism, allowing the emergence of fractal structures in the total cross section.

The impact parameter is very useful as a geometrical viewpoint of the
scattering process. The squared momentum transfer $-t=|t|$ is
replaced by its conjugate variable $b$, the transverse
distance between the colliding particles in impact parameter
space. In this space, the analytic function
$F(s,t)=\mathrm{Re}F(s,t)+i\mathrm{Im}F(s,t)$ representing the
elastic scattering is written at a fixed $s$ as %\cite{sdc2}
\begin{eqnarray}
\label{eq:3.1} F(s,t)={i}4\pi
s\int_{0}^{\infty}db\,bJ_{0}(b\sqrt{|t|})\Gamma(s,b) =i4\pi
s{\int_0}^{\infty}db\,bJ_{0}(b\sqrt{|t|})\bigl\{1-\exp[i{\chi}(s,b)]\bigr\},
\end{eqnarray}

\noindent where $J_{0}(x)$ is a zeroth order Bessel function,
$\Gamma(s,b)=\mathrm{Re}\Gamma(s,b)+i\mathrm{Im}\Gamma(s,b)=1-\exp[i{\chi}(s,b)]$
is the profile function and $\chi(s,b)$ is the eikonal written as $\chi(s,b)=\mathrm{Re}\chi(s,b)+i\mathrm{Im}\chi(s,b)$.
The unitarity condition connects the total
($\sigma_{\scriptsize{\mbox{tot}}}$), elastic
($\sigma_{\scriptsize{\mbox{el}}}$) and inelastic
($\sigma_{\scriptsize{\mbox{in}}}$) cross sections new the profile
function and can be written in $b$-representation as
\begin{eqnarray}
\label{eq:3.2}
2\mathrm{Re}\Gamma(s,b)=\bigl|\Gamma(s,b)\bigr|^2+G_{\scriptsize{\mbox{inel}}}(s,b),
\end{eqnarray}

\noindent where $G_{\scriptsize{\mbox{inel}}}(s,b)$ is the inelastic overlap function \cite{RMP-36-655-1964} and $\bigl|\Gamma(s,b)\bigr|^2$ represents the shadow contribution of the elastic channel
%\begin{subequations}
%\label{eq:3.3}
%\begin{equation}
%\label{eq:3.3a} \sigma_{\scriptsize{\mbox{tot}}}=2 \int d^{2}b\,\mathrm{Re}\Gamma(s,b),
%\end{equation}
%\begin{equation}
%\label{eq:3.3b} \sigma_{\scriptsize{\mbox{el}}}= \int d^{2}b\bigl|\Gamma(s,b)\bigr|^2,
%\end{equation}
%\begin{equation}
%\label{eq:3.3c} \sigma_{\scriptsize{\mbox{in}}}= \int d^{2}b\,G_{\scriptsize{\mbox{in}}}(s,b).
%\end{equation}
%\end{subequations}
\begin{eqnarray}
\label{eq:3.3} \sigma_{\scriptsize{\mbox{tot}}}(s)=2 \int
d^{2}b\,\mathrm{Re}\Gamma(s,b),~~~~~
\sigma_{\scriptsize{\mbox{el}}}(s)= \int
d^{2}b\bigl|\Gamma(s,b)\bigr|^2,~~~~~
\sigma_{\scriptsize{\mbox{in}}}(s)= \int
d^{2}b\,G_{\scriptsize{\mbox{inel}}}(s,b).
\end{eqnarray}

\noindent Here it is used the optical
theorem $s^{-1}\mathrm{Im}F(s,0)=\sigma_{\scriptsize{\mbox{tot}}}(s)$
\cite{Collins-book-1977,Barone-book-2002}. Also the unitarity
condition demands $\mathrm{Im}\chi(s,b)\geq 0$, implying that
$G_{\scriptsize{\mbox{inel}}}(s,b)$ represents the probability of an
elastic scattering in the 2D kinematic space $(s,b)$ and this
quantity is given by \cite{sdc2}
\begin{eqnarray}
\label{eq:3.4}
G_{\scriptsize{\mbox{inel}}}(s,b)=1-\exp\bigl[-2\mathrm{Im}\chi(s,b)\bigr]\leq
1.
\end{eqnarray}

\noindent As usual, one can introduce the opacity $\Omega(s,b)$
defined as $\Omega(s,b)=2\mathrm{Im}\chi(s,b)$. It is well-known that the opacity measures the matter density distribution inside
the incident particles. At lower energies \cite{Tevatron}, the
opacity presents a Gaussian shape not observed in LHC data, whose
indication is the growth of opacity at small $b$ as presented in
TOTEM data \cite{totem}. The inelastic profile function, on the
other hand, determines how absorptive is the interaction region
(inelastic) depending on $b$. When
$G_{\scriptsize{\mbox{inel}}}(s,b)=0$ the object is called transparent
and to $G_{\scriptsize{\mbox{inel}}}(s,b)=1$, the absorption is maximal. Theoretically the latter result, in general,
occurs at $b=0$ fm in the asymptotic condition
$s\rightarrow\infty$. Notwithstanding, in \cite{dremin_1,dremin_2} there is an approach indicating that at $b=0$ fm, the black disk limit is not achieved. Indeed, the black disk picture seems to be achieved at some critical $b_c$, near forward direction, indicating the arising of a gray area \cite{dremin_1,dremin_2}. This model was further analyzed in \cite{broniowski_arriola,alkin_martynov,anisovich_nikonov,troshin_tyurin_1,anisovich,troshin_tyurin_2,albacete_sotoontoso,arriola_broniowski} resulting in the so-called hollowness effect near the forward direction.

Using the Fourier--Bessel transform of the amplitude
(\ref{eq:3.1}) one can write the dimensionless profile function
as
\begin{eqnarray}
\label{eq:3.5} \displaystyle
\Gamma(s,b)=\frac{-i}{8\pi}\int_{0}^{\infty}d|t|J_{0}(b\sqrt{|t|})\frac{F(s,t)}{s}
\equiv \frac{-i}{8\pi}F(s,b).
\end{eqnarray}
\noindent Neglecting derivative dispersion contributions, one can suggest that the $t$-dependence is taken into account by only one real function $f(s,t)$, which is common for both real and imaginary part of $F(s,t)$. Thus, $\mathrm{Re}F(s,t)= \mathrm{Re}F(s,0)f(s,t)$ and
$\mathrm{Im}F(s,t)=\mathrm{Im}F(s,0)f(s,t)$. Also the last relations imply $f(s,0)=1$. Note that for small $|t|$ (large $b$) inside the Lehmann--Martin ellipse, $f(s,t)$ can be taken in a first approximation as the same function for the real and imaginary part of the elastic scattering amplitude, as can be viewed in \cite{avila_epj_2006}. For large $|t|$ (small $b$), $f(s,t)$ is also approximately the same for $\mathrm{Re}F(s,t)$ and $\mathrm{Im}F(s,t)$. Moreover, it should be stressed that we are mostly interested here in the large $|t|$ region (small $b$). Then
\begin{subequations}
\label{eq:3.6}
\begin{equation}
\label{eq:3.6a} \displaystyle
\mathrm{Re}\Gamma(s,b)=\frac{1}{8\pi}\frac{\mathrm{Im}F(s,0)}{s}\int_{0}^{\infty}d|t|J_{0}(b\sqrt{|t|})f(s,t)
= \frac{\sigma_{\scriptsize{\mbox{tot}}}}{8\pi}f(s,b),
\end{equation}
\begin{equation}
\label{eq:3.6b} \displaystyle
\mathrm{Im}\Gamma(s,b)=-\frac{1}{8\pi}\frac{\mathrm{Re}F(s,0)}{s}\int_{0}^{\infty}d|t|J_{0}(b\sqrt{|t|})f(s,t)
= -\rho\frac{\sigma_{\scriptsize{\mbox{tot}}}}{8\pi}f(s,b) \equiv
-\rho\mathrm{Re}\Gamma(s,b),
\end{equation}
\end{subequations}
\noindent where $\rho(s) \equiv
\mathrm{Re}F(s,0)/\mathrm{Im}F(s,0)$ is the ratio of the real
to imaginary part of the amplitude in the forward direction. It
should be noted that experimental data show $|\rho(s)| \lesssim
0.3$ (0.2) at $\sqrt{s} \gtrsim 5\,(3)$ GeV for $pp$ ($\bar{p}p$)
collisions. Therefore one can neglect the $\mathrm{Re}F(s,t)$
with respect of the $\mathrm{Im}F(s,t)$ and, as consequence, the
$\mathrm{Im}\Gamma(s,b)$ with respect to
$\mathrm{Re}\Gamma(s,b)$ at accuracy level not worse than 0.3 (0.2) in the
energy domains indicated above for $pp$ and $\bar{p}p$ collisions \cite{okorokov_int_j_mod_phys_a32_1750175_2017}.
Based on the equations (\ref{eq:3.6a}) and (\ref{eq:3.6b}) one can derive for the inelastic overlap function
\begin{eqnarray}
\label{eq:3.7}
G_{\scriptsize{\mbox{inel}}}(s,b)&=&\mathrm{Re}\Gamma(s,b)\bigl\{2-\mathrm{Re}\Gamma(s,b)[1+\rho^{2}(s)]\bigr\}
\nonumber\\
&\approx&\mathrm{Re}\Gamma(s,b)\bigl\{2-\mathrm{Re}\Gamma(s,b)\bigr\},
\end{eqnarray}
\noindent where the approximate relation is valid at accuracy
level not worse than 0.09 (0.04) in wide energy domain $\sqrt{s} \gtrsim
5\,(3)$ GeV for $pp$ ($\bar{p}p$) collisions.

It should be noted that the results obtained above are derived with the help of the most general property of quantum field theory, namely, unitarity condition and, consequently, they are model independent.

%%%%%%%%%%%%%%%%%%%%%%%%%%%%%%%%%%%%%%%%%%%%%%%%%%%%%%%%%%%%%%%%%%%%%%%%
\section{\label{bkt}The Tsallis Entropy} %sect1

Although entropy is a well-defined quantity in physics its calculation depends on the presence or not of correlations among the components of the lattice, for instance. The internal structure of the hadron grows in complexity as the energy increases, possibly showing a black disk picture as $s\rightarrow\infty$, where $s$ is the squared-energy in the center-of-mass system. This result prevents the use of the Boltzmann entropy since the correlation between the constituents of the hadron grows as $s$ increases due to the confinement potential. Accordingly, the use of the TE may furnish a better understanding of the internal structure of the hadron than the Boltzmann one.

Note that the Boltzmann entropy ($S_{B}$) assumes each part of the lattice as an independent system with a defined entropy, and the sum of all cells results in the total entropy of the system. Then, this entropy is additive, i.e., for a system composed of a countable number of subsystems $i=1,2,...,W$ each one with entropy $S_{i}$ the total entropy is given by
\begin{eqnarray}
\label{eqn1_sect2} S_{B}=\sum_{i=1}^{W} S_{i}.
\end{eqnarray}

Indeed, if the subsystems have no correlations at all or only local correlations, then $S_{B}$ is also extensive \cite{tsallis}. However, if there are correlations between the cells the Boltzmann entropy is no longer valid and the entropy of each cell cannot be computed separately and consequently, the total entropy is not the sum of each cell of the lattice. Of course, if the correlations are weak, then the Boltzmann entropy can be used as a first approximation to solve the problem. An alternative approach, however, which takes into account the correlations is provided by the TE ($S_{T}$) \cite{tsallis}. In a single system composed of two subsystems $a$ and $b$ it is written as
\begin{eqnarray}
\label{eqn2_sect2} S_{T}=S_{a} + S_{b} +(1-w)S_{a}S_{b},
\end{eqnarray}

\noindent where $w$ is the entropic index. If $w\neq 1$, then $S_T$ is non-additive as well as non-extensive. On the other hand, if $w=1$, the TE is reduced to the Boltzmann entropy one. Thus, $w$ characterizes the degree of non-extensivity of the system. It is important to stress this kind of entropy calculation is applicable when the system exhibits some long-range correlations, intrinsic fluctuations or fractal structure in phase space \cite{navarra}. Thus one notes, in the field of high energy physics the non-extensivity of the system of secondary particles can be provided by event-by-event fluctuations of some parameter that can be associated with the temperature of part of the system acting as a heat bath \cite{PRL-122-120601-2019}. Such fluctuations can lead to the violation of the strong system independence (SSI) and, consequently, to the appearance of non-extensive entropy classes, in particular, of $S_T$. Therefore, in general, the $S_T$ can be emerged in high-energy collisions due to either physical or statistical (a relatively small number of elements in the subsystem -- heat bath) reasons. An additional study and justification may be required for a conclusion about the reason for $S_T$ appearance in some interaction cases between particles and nuclei. The continuous form of the TE is given by \cite{capek_book_2005}
\begin{eqnarray}
\label{eqn3_sect2}
S_{T}(p,w)=\frac{1}{w-1}\left(1-\int_{-\infty}^{\infty}\bigl[p\,(x)\bigr]^{w}dx\right),
\end{eqnarray}

\noindent where $p\,(x)$ is the probability density function. The existence of fractal structures in the momentum and in the energy space for both $pp$ and $\bar{p}p$ collisions may allow the use of the TE to study how these fractal dimensions can contribute to the understanding of the collision process. Additionally, if one looks to the entropy as the information that may be gained observing a physical quantity depending on ($s,b$) in the impact parameter space, then $b$ may be used as a measure of this information and how it acts on the black disk picture.

%%%%%%%%%%%%%
\subsection{Tsallis Entropy in the Impact Parameter Space}

In order to use the TE in the impact parameter space one assumes the following assumptions:
\begin{itemize}
\item[1.]{the probability density function is calculated inside a disk of radius $b$, the impact parameter;}
\item[2.]{the entropic index $w$ can be replaced by a single real evaluated function $w=w(s/s_{c})\geq 0$, where $s_{c}$ is the critical point associated to the phase transition occurring in the total cross section \cite{borcsik_mod_phys_lett_a31_1650066_2016}.}
\end{itemize}

The first assumption is necessary to establish an ordering in the elastic scattering process considering the growth of the black disk picture as $s$ increases. In the impact parameter space, all functions are $b$-dependent (fixed-$s$) and at each value of $s$ ($s_{1}<s_{2}<...<s_{n}$), the description of the elastic scattering acquires a black disk behavior as shown by the profile function and the inelastic overlap function \cite{sdc2}, i.e. the hadron radius grows with the energy. Then, at each $s_{i}$ there is a maximum value to $\mathrm{Re}\Gamma(s_{i},b=0)$, for instance, with an effective maximum range (or radius) $b_{i}$ ($i=1,2,...,n$). Hence, one computes the amount of information inside the disk of radius $b_{i}$.

The second assumption is related to the critical energy value $\sqrt{s_c}$ where the phase transition takes place. The energy range where $\sigma_{\scriptsize{\mbox{tot}}}(s)$ change its curvature is $\sqrt{s}\sim 10 - 30$ GeV, depending on the dataset considered ($pp$ or $\bar{p}p$). This critical energy value may represent a phase transition due to some change in the arrangement of the hadron internal constituents. According to \cite{borcsik_mod_phys_lett_a31_1650066_2016}, there are two fractal dimensions in the total cross section experimental dataset to $pp$ and $\bar{p}p$. The non-extensivity of the system is given by $s/s_{c}$ and it represents a measure of the correlations among the internal constituents of $pp$ and $\bar{p}p$. The phase transition occurring in the total cross section at $s_c$ represents a break in the symmetry predicted in the seminal paper \cite{cheng_wu_phys_rev_lett_24_145_1970} and, before that, in the \cite{heisenberg_z_phys_133_65_1952}. Therefore, the entropic index can be chosen in the form $w=(s/s_{c})^{\alpha}$. It should be stressed that the entropic index growth with collision energy is confirmed by numerous particle production studies, in some strong interaction processes \cite{PLB-723-351-2013,EPJC-74-2785-2014,AHEP-2016-9632126-2016,EPJA-53-102-2017}.
The information about $\alpha$ is limited and, for instance, the study of negative charged pions in $pp$ indicates the $\alpha \sim 0.007$ for the fitting function $w \propto s^{\alpha}$ \cite{EPJC-74-2785-2014}. However, the value of the $\alpha$ parameter can depend on the particle species, collision type, etc. There are no theoretical and /or experimental restrictions for value of this parameter. In the present energy range, the real prediction of $\alpha$ behavior cannot be settled down. Moreover, one assumes $\alpha>0$ and, then, its value does not affect the interpretation of the entropy derived here. Thus in the present paper, $\alpha=1$ is used for simplicity and without loss of generality.

These assumptions can provide a way to calculate the entropy generated due a phase transition in the total cross section as $s$ increases \cite{borcsik_mod_phys_lett_a31_1650066_2016}. In order to do that, the integral in (\ref{eqn3_sect2}) is rewritten assuming the collision event inside a disk of radius $b>0$ and an entropic index $s/s_{c}$
\begin{eqnarray}
\label{eqnasq_sect4}1-\int_{0}^{\infty}\bigl[p\,(b')\bigr]^{s/s_{c}}db'
=1-\int_{0}^{b}\bigl[p\,(b')\bigr]^{s/s_{c}}db'=m[1-nG_{\scriptsize{\mbox{inel}}}(s,b)], %\equiv m(1-nG),
\end{eqnarray}

\noindent where $m, n$ are dimensionless real free parameters (to prevent entropy complex values). Although the concept of complex entropy can be well-defined in information theory \cite{rotundo_eur_phys_j_b86_169_2013}, it will be developed in particle scattering elsewhere. Since $\displaystyle P(s,b)=\int_{0}^{b}\bigl[p\,(b')\bigr]^{s/s_{c}}db'$ is a probability and $0 \leq P(s,b) \leq 1$, then $0 \leq n \leq G^{-1}_{\scriptsize{\mbox{inel}}}(s,b)$.
It is important to stress that the integration upper limit acts as a cutoff in the impact parameter space. This cannot be viewed as a method limitation since it is expected that $b\rightarrow\infty$, all $b$-dependent functions vanishes. Hence, contributions above some $b$ can be neglected assuming an effective range of interaction. Therefore, the TE can be written in terms of the inelastic overlap function as ($k \equiv s/s_{c}-1$)
\begin{eqnarray}
\label{eqnsw_sect4}
S_{T}(s,b)&=&mk^{-1}\bigl[1-nG_{\scriptsize{\mbox{inel}}}(s,b)\bigr]
=mk^{-1}\bigl(1-n\mathrm{Re}\Gamma(s,b)\bigl\{2-\mathrm{Re}\Gamma(s,b)[1+\rho^{2}(s)]\bigr\}
\bigr) \nonumber\\
&\approx&mk^{-1}\bigl(1-n\mathrm{Re}\Gamma(s,b)\bigl\{2-\mathrm{Re}\Gamma(s,b)\bigr\}\bigr).
\end{eqnarray}

\noindent The above relations can be rewritten as follows
\begin{eqnarray}
\label{eqnsw_sect4.a}
S_{T}(s,b)=mk^{-1}\bigl[\mathrm{Re}\Gamma(s,b)-X_{1}\bigr]
\bigl[\mathrm{Re}\Gamma(s,b)-X_{2}\bigr],
\end{eqnarray}

\noindent where $X_{i}=[1+\rho^{2}(s)]^{-1}\bigl(1 \pm \sqrt{1-[1+\rho^{2}(s)]/n}\bigr) \approx 1 \pm \sqrt{1-1/n}$, $i=1,2$. Note that $m$ and $n$ rules as scales for the problem and does not alter the physical interpretation of any result obtained below, unless $0 \leq n < 1$, where $\forall\,i=1, 2: X_{i}$. Thus, for the sake of simplicity, one adopts $n=1$ and $m=1$ and the TE assumes its symmetric form in the limiting case $\rho \to 0$
\begin{eqnarray}
\label{dremin_4_2}
S_{T}(s,b)&=&k^{-1}\bigl[1-G_{\scriptsize{\mbox{inel}}}(s,b)\bigr]
=k^{-1}\bigl(1-\mathrm{Re}\Gamma(s,b)\bigl\{2-\mathrm{Re}\Gamma(s,b)[1+\rho^{2}(s)]\bigr\}
\bigr) \nonumber\\
&\approx&k^{-1}\bigl[1-\mathrm{Re}\Gamma(s,b)\bigr]^{2}.
\end{eqnarray}

The above result furnishes a measure of the entropy in the impact parameter space through the use of the inelastic overlap function. It is known that $G_{\scriptsize{\mbox{inel}}}(s,b)$ can be evaluated with the help of some model--dependent technique. Therefore the $S_{T}(s,b)$ defined by the relations (\ref{eqnsw_sect4}) -- (\ref{dremin_4_2}) is model--dependent in general. The signature (positive or negative) of the TE reveals an $s$-dependence analyzed as follows. Considering $s<s_{c}$, the signature of $S_{T}$ is negative and it can be interpreted as the system using the energy of the beam to self-organize or maintain its internal structure (quarks and gluons). Then, as $s$ increases the entropy tends to a maximum by negative values, i.e. the system tends to achieve the maximum of its self-organization as well. Moreover, using a simple fitting model written as
\begin{eqnarray}\label{eq:fit_1}
\sigma_{\scriptsize{\mbox{tot}}}(s)=\gamma_1\ln(s/s_c)^{\gamma_2}
\end{eqnarray}

\noindent where $\gamma_1$ and $\gamma_2$ are free fit parameters, being $\gamma_2$ the Hausdorff--Besicovitch fractal dimension, a novel interpretation for the total cross section was proposed in \cite{borcsik_mod_phys_lett_a31_1650066_2016} where the fractal dimension in this energy range is negative and different to $pp$ and $\bar{p}p$, producing distinct patterns to $G_{\scriptsize{\mbox{inel}}}(s,b)$. Hence, the measure of the emptiness of $pp$ and $\bar{p}p$ total cross section results in different values to $S_{T}$.

The negative entropy and the negative fractal dimensions imply the constituents, the internal arrangement to $pp$ and $\bar{p}p$ are unlike, i.e. the quark-quark and quark-gluon arrangement of the proton are different of the antiquark-antiquark and antiquark-gluon arrangement of the antiproton. In the popular picture, one says that the odderon distinguishes particle from antiparticle.

When the energy $s$ grows and go through the transition point $s_{c}$, the TE turns positive and stands for the system growing disorder. As obtained in \cite{borcsik_mod_phys_lett_a31_1650066_2016}, the total cross section to $pp$ and $\bar{p}p$ possesses positive fractal dimensions to $s>s_{c}$ and both tends to the same value as $s\rightarrow\infty$. Both results have shown that at high energies the arrangement of the internal constituents to $pp$ and $\bar{p}p$ tends to the same behavior. Then, the pomeron does not distinguish particle from antiparticle.

The different internal arrangement of proton and antiproton may absorb the incoming energy by distinct mechanisms. As pointed out \cite{borcsik_mod_phys_lett_a31_1650066_2016}, the negative fractal dimension represents the emptiness of the hadron internal arrangement and the total cross section is a measure of that. The internal arrangement of quarks and gluons at lower energies in the proton picture is less empty than the arrangement of the antiquarks and gluons inside the antiproton, as can be viewed in the total cross section experimental dataset for $s<s_{c}$.

At the transition point, $(s=s_{c})$ the fractal dimension to $pp$ and $\bar{p}p$ total cross section is null and the system achieves its maximum capability to convert the absorbed energy in order. This point may indicate the first saturation point in $pp$ and $\bar{p}p$ total cross section dataset. It is interesting to note that $pp$ and $\bar{p}p$ total cross section tends to the same saturation point $s_c$, possibly indicating this value as a universal character of total cross sections.

Above the critical point, $(s>s_{c})$, the internal constituents of $pp$ and $\bar{p}p$ achieve degrees of freedom previously blocked by using the energy coming from the beam converted in thermal agitation resulting in the rise of the total cross-section as $s$ increases.

The above scenarios introduced by the TE and by the fractal dimension concept result in the question of how occurs the filling up mechanism responsible by the black disk behavior of $pp$ and $\bar{p}p$ as $s\rightarrow\infty$. A possible answer is given as follows. As well-known, in QCD the confinement of quarks and gluons prevent its freedom below the Hagedorn temperature \cite{hagedorn_nuovo_cim_supp_3_147_1965}, where the hadrons are no longer stable. However, the increasing energy of the scattering imply in the enhanced of the thermal bath at each particle is subject. This energy is then transferred to the internal constituents of the proton and antiproton by a heat transport mechanism.

The zero entropy state can be established when at $nG_{\scriptsize{\mbox{inel}}}(s,b)=1$ for some particular $(s,b)$. As well-known, zero entropy occurs when a system achieves its ground-state (or its maximum self-organization state). Thus, at this point, the physical state of the system is completely known (the ways one can arrange its internal configuration is exactly one). The general belief is that $nG_{\scriptsize{\mbox{inel}}}(s,b)=1$ is achieved in the asymptotic limit $s\rightarrow\infty$ and at $b=0$. Then, from some $s$ sufficiently high the energy of both $pp$ and $\bar{p}p$ possess the same behavior at $b=0$. However, there exist some models indicating this result may be achieved at some $b\neq 0$ \cite{dremin,broniowski_arriola}. The implication of that is the appearance of a gray area in the inelastic overlap function near $b=0$.

It is interesting to note that the first equation in the chain (\ref{eqnsw_sect4}) can also be written assuming only as the first order of the logarithm expansion below
\begin{eqnarray}
\label{eqnswmodif_sect4} S_{T}(s,b)=-mk^{-1}\ln \bigl[nG_{\scriptsize{\mbox{inel}}}(s,b)\bigr],
\end{eqnarray}

\noindent implying higher orders are corrections for equation (\ref{eqnsw_sect4}). In addition, if the logarithm of the inelastic overlap function is connected to the pair spatial separation of the constituents of the hadron, then it can represent the interaction of topological defects \cite{berezinskii_zh_eksp_teor_fiz_59_907_1970,kosterliz_j_phys_c6_1181_1973} inside the hadron. As well-known, in the BKT phase transition, the entropy depends on the logarithm of the spatial pair separation of vortices. On the other hand, it has been shown that the correct description of the inelastic overlap function needs at least two Gaussian \cite{fagundes}. If one associate each Gaussian to a particular location inside the hadron, then the TE given by equation (\ref{eqnswmodif_sect4}) may be interpreted as a BKT-like phase transition occurring inside the hadron at $s=s_c$, being $s_c$ the critical squared energy value where the total cross section experimental dataset change its curvature.

\section{\label{comp}Basic Application: General Form for the Overlap Function}

In this section one focus on recent inelastic overlap function
models comparing the results by using $S_{T}$. These
comparisons may furnish a better understanding of how entropy is
released in each model. There is a wide set of models for
nucleon-nucleon elastic scattering and, therefore, it seems reasonable
to discuss only those based on the most general and basic
statements of the Axiomatic Quantum Field Theory (AQFT). In the present paper, the (a) unitarity condition and (b) asymptotic theorems are the basic ground.

\subsection{Non-central collisions ($b \ne 0$)}\label{subsec:4.a}

The most general and well-established experimental result for elastic
scattering is the fast decreasing of the differential cross
section ($d\sigma / dq^{2}$) with the increasing $|t| \simeq
q^{2}$ in the diffraction peak. As a first approximation, the
$d\sigma / dq^{2}$ shows an exponential growth with the slope $B(s)$ at
$q^{2}$ under consideration. Thus, one writes for the $t$-dependent
part of scattering amplitude $f(s,t)=\exp[B(s)|t|/2]$ and
\begin{eqnarray}
\label{eq:4.a.1} \displaystyle
f(s,b)=\int_{0}^{\infty}d|t|J_{0}(b\sqrt{|t|})\exp\biggl[\frac{B(s)|t|}{2}\biggl]
\approx
2\int_{0}^{\infty}dq\,qJ_{0}(bq)\exp\biggl[-\frac{B(s)q^{2}}{2}\biggl]
= \frac{2}{B(s)}\exp\biggl[-\frac{b^{2}}{2B(s)}\biggr].
\end{eqnarray}
\noindent Then
\begin{subequations}
\label{eq:4.a.2}
\begin{equation}
\label{eq:4.a.2a}
\mathrm{Re}\Gamma(s,b)=\bigl[\sigma_{\scriptsize{\mbox{tot}}}/4\pi
B(s)\bigr]\exp\bigl[-b^{2}/2B(s)\bigr] \equiv
\zeta(s)\exp\bigl[-b^{2}/2B(s)\bigr],
\end{equation}
\begin{equation}
\label{eq:4.a.2b}
\mathrm{Im}\Gamma(s,b)=\bigl[-\rho\sigma_{\scriptsize{\mbox{tot}}}/4\pi
B(s)\bigr]\exp\bigl[-b^{2}/2B(s)\bigr] \equiv
-\rho(s)\zeta(s)\exp\bigl[-b^{2}/2B(s)\bigr],
\end{equation}
\end{subequations}
\noindent where $\zeta$ is the parameter defined as following
\begin{eqnarray}
\label{eq:4.a.3} %\zeta(s)=\sigma_{\scriptsize{\mbox{tot}}}(s) / 4\pi B(s) \approx 4\sigma_{\scriptsize{\mbox{el}}}(s) / \sigma_{\scriptsize{\mbox{tot}}}(s).
\zeta(s)=\frac{\textstyle
\sigma_{\scriptsize{\mbox{tot}}}(s)}{\textstyle 4\pi B(s)}=
\frac{\textstyle 4\sigma_{\scriptsize{\mbox{el}}}(s)}{\textstyle
[1+\rho^{2}(s)]\sigma_{\scriptsize{\mbox{tot}}}(s)} \approx
\frac{\textstyle 4\sigma_{\scriptsize{\mbox{el}}}(s)}{\textstyle
\sigma_{\scriptsize{\mbox{tot}}}(s)}.
\end{eqnarray}
\noindent Note that at $\zeta=1$ the
$G_{\scriptsize{\mbox{inel}}}(s,b)$ represent a black disk and for
$\zeta\neq 1$ the inelastic overlap function diminishes. Moreover,
the position of the maximum
$b_{\scriptsize{\mbox{max}}}^{2}=2B\ln\zeta$ with the full absorption
$G_{\scriptsize{\mbox{inel}}}(s,b_{\scriptsize{\mbox{max}}})=1$
depends on $B(s)$ and $\zeta(s)$.

Taking into account (\ref{eq:3.7}) and (\ref{eqnsw_sect4}) one
can deduce the final expressions for both the inelastic overlap function
and the TE, respectively, within a general phenomenological way for the scattering amplitude
\begin{eqnarray}
\label{eq:4.a.4}
G_{\scriptsize{\mbox{inel}}}(s,b)&=&\zeta(s)\exp\bigl[-b^{2}/2B(s)\bigr]\bigl\{2-\zeta(s)\exp\bigl[-b^{2}/2B(s)\bigr]
[1+\rho^{2}(s)]\bigr\}
\nonumber\\
&\approx&\zeta(s)\exp\bigl[-b^{2}/2B(s)\bigr]\bigl\{2-\zeta(s)\exp\bigl[-b^{2}/2B(s)\bigr]\bigr\},
\end{eqnarray}
\begin{eqnarray}
\label{eq:4.a.5}
S_{T}(s,b)&=&mk^{-1}\bigl(1-n\zeta(s)\exp\bigl[-b^{2}/2B(s)\bigr]\bigl\{2-\zeta(s)\exp\bigl[-b^{2}/2B(s)\bigr]
[1+\rho^{2}(s)]\bigr\}\bigr) \nonumber\\
&\approx&mk^{-1}\bigl(1-n\zeta(s)\exp\bigl[-b^{2}/2B(s)\bigr]\bigl\{2-\zeta(s)\exp\bigl[-b^{2}/2B(s)\bigr]\bigr\}).
\end{eqnarray}

%For strong interaction processes, the temperature in the above model independent approach for the $G_{\scriptsize{\mbox{inel}}}(s,b)$ can be performed using (\ref{eqnsw_sect4.a}), resulting the general case
%\begin{eqnarray}
%\displaystyle
%\label{temp_1} T(s,b)= \frac{k\bigl(-4\alpha_s(\eta s)/3\varepsilon_{2}b+\kappa \varepsilon_{2} b\bigr)}{m\bigl[\mathrm{Re}\Gamma(s,b)-X_{1}\bigr]
%\bigl[\mathrm{Re}\Gamma(s,b)-X_{2}\bigr)]},
%\end{eqnarray}

%\noindent where the symmetric form is written as
%\begin{eqnarray}
%\displaystyle
%\label{temp_2} T(s,b)= \frac{k\bigl(-4\alpha_s(\eta s)/3\varepsilon_{2}b+\kappa \varepsilon_{2} b\bigr)}{\bigl[\mathrm{Re}\Gamma(s,b)-1\bigr]^{2}}.
%\end{eqnarray}

%The above result agrees with the analyzes performed in the preceding section showing the picture of the smoothed torus when $s<s_c$ and a disk-like if $s>s_c$ tending to a point-like particle in the asymptotic energy limit.

\subsection{Central collisions ($b=0$)}\label{subsec:4.b}

The general equations (\ref{eq:4.a.4}) and (\ref{eq:4.a.5}) are obtained assuming the most general phenomenological view for the differential cross section
$d\sigma/dq^{2}=\bigl[\mathrm{Im}F(s,0)/4\sqrt{\pi}s\bigr]^{2}\exp\bigl[-B(s)q^{2}/2\bigr]$. However, the concrete form for the energy dependence of both the $G_{\scriptsize{\mbox{inel}}}(s,b)$ and the $S_{T}(s,b)$ is driven by the corresponding dependence for scattering parameters. The exact relations in (\ref{eq:4.a.4}), (\ref{eq:4.a.5}) are defined by energy dependencies for global scattering parameters $\sigma_{\scriptsize{\mbox{tot}}}$, $\rho$ and for the slope $B$, while the corresponding approximate relations depend on the ratio $R_{\scriptsize{\mbox{e/t}}} = \sigma_{\scriptsize{\mbox{el}}}/\sigma_{\scriptsize{\mbox{tot}}}$ and $B$. The forward condition $b=0$ allows the exclusion of the dependence on $B(s)$. In this specific case, $\mathrm{Re}\Gamma(s,0)=\zeta(s)$ and $\mathrm{Im}\Gamma(s,0)=-\rho(s)\zeta(s)$. Considering exactly central collision with $b=0$ one obtains from the general relations (\ref{eq:4.a.4}), (\ref{eq:4.a.5}) the following results
\begin{eqnarray}
\label{eq:4.b.1}
G_{\scriptsize{\mbox{inel}}}(s,0)=\zeta(s)\bigl\{2-\zeta(s)[1+\rho^{2}(s)]\bigr\}
\approx \zeta(s)\bigl\{2-\zeta(s)\bigr\},
\end{eqnarray}
\begin{eqnarray}
\label{eq:4.b.2}
S_{T}(s,0)&=&mk^{-1}\bigl\{1-n\zeta(s)\bigl[2-\zeta(s)(1+\rho^{2}(s))\bigr]\bigr\}
\approx mk^{-1}\bigl\{1-n\zeta(s)\bigl[2-\zeta(s)\bigr]\bigr\}.
\end{eqnarray}

The symmetric form for the TE in central collisions
\begin{eqnarray}
\label{eq:4.b.2a}
S_{T}(s,0)=k^{-1}\bigl[1-\zeta(s)\bigr]^{2}
\end{eqnarray}

\noindent derived from (\ref{eq:4.b.2}) can also be viewed as the first order approximation of the logarithm series $S_{T}(s,0)=k^{-1}\ln^2\zeta$. Thus, for exactly central $pp$ and $\bar{p}p$ collisions, only energy dependence for $R_{\scriptsize{\mbox{e/t}}}$ remains, which varies in different models. Detailed analysis of this dependence for $pp$ and $\bar{p}p$ scattering as well as for joined sample for these collisions\footnote{Below for brevity the joined sample is also called the ensemble for nucleon-nucleon scattering.} is made in \cite{okorokov-arXiv-1805.10514} with the help of the fitting of experimental data by an empirically chosen function.

In general, the asymptotic value $\left.\zeta(s)\right|_{s \to \infty}$ varies from one approach to another due to model-dependent value of $R_{\scriptsize{\mbox{e/t}}}$ for $s \to \infty$. The result from \cite{okorokov-arXiv-1805.10514}, obtained with the help of asymptotic theorems and assumptions for the properties of scattering amplitude for binary process $1+2 \to 3+4$ within AQFT, assumes that $\left.\zeta(s)\right|_{s \to \infty} \to 3$. The approach of partonic disks \cite{PU-58-963-2015} provides some faster growth of the ratio of elastic to total cross section, which leads to the $R_{\scriptsize{\mbox{e/t}}}(s) \to 1$ for $s \to \infty$ and, consequently, $\left.\zeta(s)\right|_{s \to \infty} \to 4$. It should be noted that $\left.G_{\scriptsize{\mbox{inel}}}(s,0)\right|_{s \to \infty} < 0$ in accordance with the (\ref{eq:4.b.1}) within models with $\left.\zeta(s)\right|_{s \to \infty} > 2$. On the other hand, if the total cross section in the asymptotic energy domain is half one has usually today, i.e. is bounded by a modified Froissart--Martin limit $\left.\sigma_{\scriptsize{\mbox{tot}}}(s)\right|_{s \to \infty} < (\pi/2m_{\pi}^{2})\ln^{2}\varepsilon$ \cite{martin_phys_rev_d80_065013_2009}, then the inelastic cross section bounded by $(\pi/4m_{\pi}^{2})\ln^{2}\varepsilon$ is two times smaller, where $m_{\pi}$ is the pion mass \cite{PDG-PRD-98-030001-2018}, $\varepsilon \equiv s/s_{0}$ and $s_{0}=1$ GeV$^{2}$. Consequently, $R_{\scriptsize{\mbox{e/t}}}(s) \to 1/2$ for $s \to \infty$. Furthermore, the harder boundary result $\left.R_{\scriptsize{\mbox{e/t}}}(s)\right|_{s \to \infty} < 1/2$ can be obtained if is accepted that the elastic cross section cannot be larger than the inelastic cross section ($\sigma_{\scriptsize{\mbox{inel}}}$), the limiting case being an expanding black disk \cite{PRD-91-076006-2015}. This assumption allows the restoration of $\zeta(s)$ into the interval (0,2). However, if the modified Froissart--Martin limit \cite{martin_phys_rev_d80_065013_2009} is overcomed and / or $\sigma_{\scriptsize{\mbox{el}}} > \sigma_{\scriptsize{\mbox{inel}}}$ can be for $s \to \infty$, then in such approaches the identification of the inelastic overlap function with some probability requires additional study and justification for asymptotic energies\footnote{Usually phenomenological models consider the energy domain no wider than $\sqrt{s} \gtrsim 3-5$ GeV excluding the narrow range on $s$ close to the low-energy boundary $s_{\scriptsize{\mbox{l.b.}}} \equiv 4m_{p}^{2}$ for the interactions ($pp$, $\bar{p}p$) under discussion in which $R_{\scriptsize{\mbox{e/t}}}$ reaches large values in $pp$, where $m_{p}$ is the proton mass \cite{PDG-PRD-98-030001-2018}. Moreover, this low-energy range is excluded in the present analysis due to above condition $|\rho(s)| \lesssim 0.3$.}.

The experimental database for $R_{\scriptsize{\mbox{e/t}}}(s)$ and fit results for this quantity are taken from \cite{okorokov-arXiv-1805.10514} and are used in order to evaluate the energy dependencies for both $G_{\scriptsize{\mbox{inel}}}(s,0)$ and $S_{T}(s,0)$ in $pp$, $\bar{p}p$ elastic scattering\footnote{In the paper total errors are
used for estimations based on the experimental points for $R_{\scriptsize{\mbox{e/t}}}(s)$, unless otherwise specified. The total
error is calculated as addition of systematic and statistical
uncertainties in quadrature \cite{PDG-PRD-98-030001-2018}.}. Among analytic functions suggested in \cite{okorokov-arXiv-1805.10514}, the approximation with the power law term $\propto \varepsilon^{-\beta}$ leads to a slightly better description of experimental data for $R_{\scriptsize{\mbox{e/t}}}$ than the function with the term $\propto \ln^{-\gamma}\varepsilon$ at low boundaries for fitted intervals on energy $\sqrt{s_{\scriptsize{\mbox{min}}}} \geq 3$ GeV. In accordance with the discussion above, the approximate relations in (\ref{eq:4.b.1}) and (\ref{eq:4.b.2}) are valid for  $\sqrt{s} \gtrsim
5\,(3)$ GeV for $pp$ ($\bar{p}p$) collisions at accuracy level not worse than 0.09 (0.04). Consequently, the following analytic function is considered for $R_{\scriptsize{\mbox{e/t}}}(s)$:
\begin{eqnarray}
R_{\scriptsize{\mbox{e/t}}}(s)&=& a_{1}+a_{2}\ln^{a_{3}}
\varepsilon +a_{4}\varepsilon^{-a_{5}}, \label{eq:4.b.3}
\end{eqnarray}

\noindent where free parameters $a_{i}$, $i=1-5$ depended on range of the fit, i.e. on the low boundary for the energy interval $s \geq
s_{\scriptsize{\mbox{min}}}$ \cite{okorokov-arXiv-1805.10514}. Moreover, the results from \cite{okorokov-arXiv-1805.10514} allow the comparison with other phenomenological approaches: the smooth curves for $G_{\scriptsize{\mbox{inel}}}(s,0)$ and $S_{T}(s,0)$ are obtained by using the $R_{\scriptsize{\mbox{e/t}}}(s)$ estimated as the ratio of approximation for $\sigma_{\footnotesize\mbox{el}}$ from \cite{TOTEM-arXiv-1712.06153} considering "standard" functions for $\sigma_{\footnotesize\mbox{tot}}$ in $pp$ and $\bar{p}p$ reactions from \cite{PDG-PRD-98-030001-2018}.

\begin{figure}
\centering
\includegraphics[width=17.0cm,height=14.0cm]{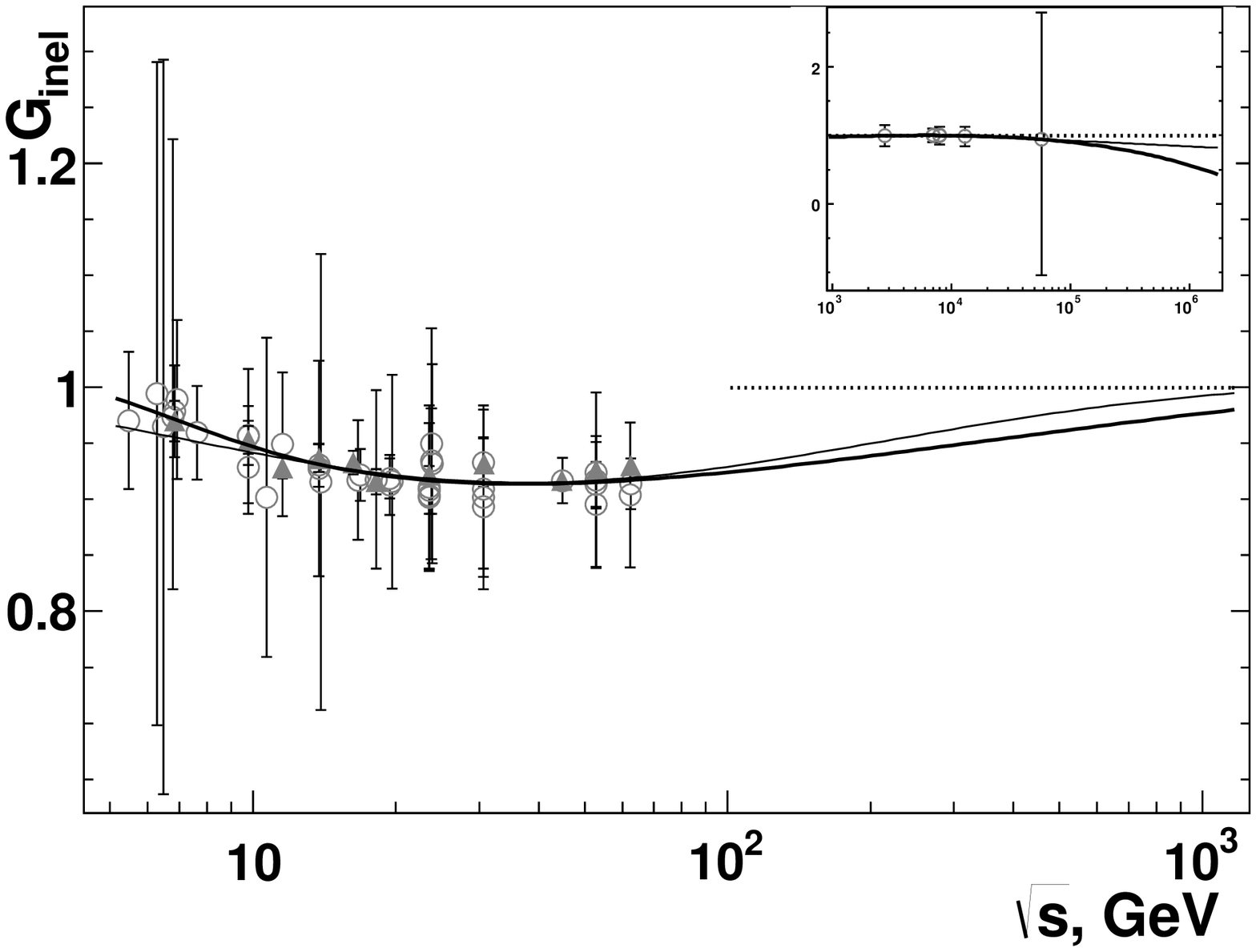}
\caption{Energy dependence for the inelastic overlap function in
central $pp$ collisions $G_{\scriptsize{\mbox{inel}}}(s,0)$.
Experimental estimations obtained within present work are shown by
open points, the solid symbols correspond to the results from
\cite{PRD-14-3092-1976,NPB-166-301-1980}, curves are evaluated
with help of the smooth dependencies for
$R_{\scriptsize{\mbox{e/t}}}^{pp}$ from
\cite{okorokov-arXiv-1805.10514}. Solid curve corresponds to the
results of the fitting of experimental
$R_{\scriptsize{\mbox{e/t}}}^{pp}(s)$ by (\ref{eq:4.b.3}) at
$\sqrt{s_{\scriptsize{\mbox{min}}}}=5$ GeV and dotted line -- by
constant at $\sqrt{s_{\scriptsize{\mbox{min}}}}=100$ GeV (see
detailed description in the text). The thin solid line is obtained
on basis of the ratio of the approximation for
$\sigma_{\scriptsize{\mbox{el}}}(s)$ from
\cite{TOTEM-arXiv-1712.06153} to the analytic function for
$\sigma_{\scriptsize{\mbox{tot}}}^{pp}(s)$ from
\cite{PDG-PRD-98-030001-2018}. Inner panel: experimental
estimations and curves for the energy domain $\sqrt{s} > 1$ TeV.}
\label{fig:1}
\end{figure}

\begin{figure}
\centering
\includegraphics[width=17.0cm,height=14.0cm]{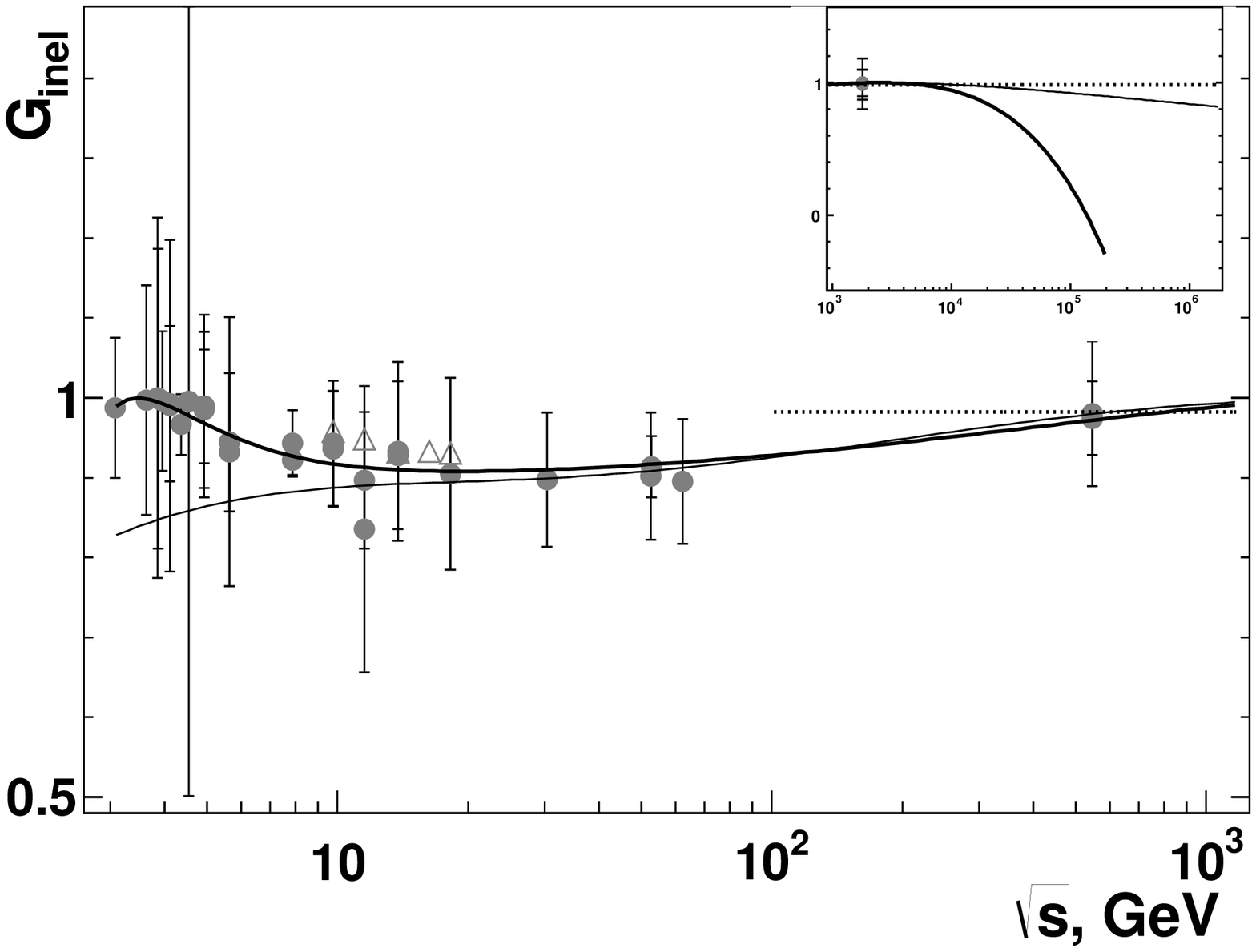}
\caption{Energy dependence for the inelastic overlap function in
central $\bar{p}p$ collisions $G_{\scriptsize{\mbox{inel}}}(s,0)$.
Experimental estimations obtained within present work are shown by
solid points, the open symbols correspond to the results from
\cite{PRD-14-3092-1976}, curves are evaluated with help of the
smooth dependencies for $R_{\scriptsize{\mbox{e/t}}}^{\bar{p}p}$
from \cite{okorokov-arXiv-1805.10514}. Solid curve corresponds to
the results of the fitting of experimental
$R_{\scriptsize{\mbox{e/t}}}^{\bar{p}p}(s)$ by (\ref{eq:4.b.3}) at
$\sqrt{s_{\scriptsize{\mbox{min}}}}=3$ GeV and dotted line -- by
constant at $\sqrt{s_{\scriptsize{\mbox{min}}}}=100$ GeV (see
detailed description in the text). The thin solid line is obtained
on basis of the ratio of the approximation for
$\sigma_{\scriptsize{\mbox{el}}}(s)$ from
\cite{TOTEM-arXiv-1712.06153} to the analytic function for
$\sigma_{\scriptsize{\mbox{tot}}}^{\bar{p}p}(s)$ from
\cite{PDG-PRD-98-030001-2018}. Inner panel: experimental
estimations and curves for the energy domain $\sqrt{s} > 1$ TeV.}
\label{fig:2}
\end{figure}

Figs. \ref{fig:1} and \ref{fig:2} show the energy dependence for
the inelastic overlap function in central $pp$ and $\bar{p}p$
collisions, respectively. Estimations for
$G_{\scriptsize{\mbox{inel}}}(s,0)$, deduced with the help of the
experimental values for $R_{\scriptsize{\mbox{e/t}}}(s)$ within
present work, are shown by points. Solid triangles in Fig.
\ref{fig:1} are from \cite{PRD-14-3092-1976,NPB-166-301-1980},
results for $G_{\scriptsize{\mbox{inel}}}(s,0)$ obtained for
$\bar{p}p$ in \cite{PRD-14-3092-1976} are shown by open symbols in
Fig. \ref{fig:2}. Smooth curves obtained in
\cite{okorokov-arXiv-1805.10514} are also re-calculated for
$G_{\scriptsize{\mbox{inel}}}(s,0)$. As seen, estimations for
$G_{\scriptsize{\mbox{inel}}}(s,0)$ deduced with the help of the ratio
of the elastic to total cross sections, agree quite well with the
results obtained by another techniques in
\cite{PRD-14-3092-1976,NPB-166-301-1980} at corresponding
$\sqrt{s}$ for $pp$ (Fig. \ref{fig:1}) and $\bar{p}p$ (Fig.
\ref{fig:2}) collisions. Such agreement confirms the validity of
the approach used here for the production of the energy dependence of
inelastic overlap function in central $pp$ and $\bar{p}p$
collisions.

In general, the experimental estimations for
$G_{\scriptsize{\mbox{inel}}}(s,0)$ deduced considering the result (\ref{eq:4.b.1}), are featured by large errors which turns difficult to obtain unambiguous physical conclusions. Considering
large errors, one can suppose that $pp$ (Fig. \ref{fig:1}) and $\bar{p}p$ scattering (Fig. \ref{fig:2}) are close to the
black disk picture for $\sqrt{s} \lesssim 10$ GeV. Then, trend is
seen for some decreasing of $G_{\scriptsize{\mbox{inel}}}(s,0)$ up
to the highest Intersecting Storage Rings (ISR) energy
$\sqrt{s} \approx 63$ GeV. There are gaps without experimental
data for both $pp$ and $\bar{p}p$, especially large for the first
case. The importance of new experimental data from Relativistic
Heavy Ion Collider (RHIC) at the interval $\sqrt{s} \sim 0.1 - 0.5$ TeV and
from low-energy Large Hadron Collider (LHC) mode for $\sqrt{s} \sim
1.0$ TeV is mentioned elsewhere \cite{okorokov-arXiv-1805.10514}.
Experimental estimations for $pp$ for $\sqrt{s} > 1$ TeV agree
quite well with the black disk picture (Fig. \ref{fig:1}, inner panel)
and this statement is valid for $\bar{p}p$ starting with
$\sqrt{s}=546$ GeV (Fig. \ref{fig:2}). As seen in Figs.
\ref{fig:1}, \ref{fig:2}, the smooth curves obtained within various
models describe reasonably the experimental estimations for the
$G_{\scriptsize{\mbox{inel}}}(s,0)$ in whole available energy
range $\sqrt{s} \geq 5\,(3)$ GeV for $pp$ ($\bar{p}p$) collisions.
This is expected due to corresponding results for
$R_{\scriptsize{\mbox{e/t}}}(s)$ \cite{okorokov-arXiv-1805.10514}
for the estimations of $G_{\scriptsize{\mbox{inel}}}(s,0)$, deduced
within the present work. Moreover, the smooth curves agree reasonably with the estimations obtained by another technique in
\cite{PRD-14-3092-1976,NPB-166-301-1980} for both $pp$ (Fig.
\ref{fig:1}) and $\bar{p}p$ (Fig. \ref{fig:2}) collisions with
some underestimation for the last case. The constant
$G_{\scriptsize{\mbox{inel}}}(s,0) \approx 1.0$ agrees with
high-energy $pp$ and $\bar{p}p$ experimental estimations.
Empirical curve based on (\ref{eq:4.b.3}) is close to the one
obtained with parameterizations from
\cite{PDG-PRD-98-030001-2018,TOTEM-arXiv-1712.06153} for $\sqrt{s}
\leq 1$ TeV for both $pp$ (Fig. \ref{fig:1}) and $\bar{p}p$ (Fig.
\ref{fig:2}) collisions. In the last case some discrepancy is seen
for $\sqrt{s} \lesssim 10$ GeV, which can be explained by the fact
that, strictly speaking,  the parametrization for
$\sigma_{\footnotesize\mbox{el}}$ from
\cite{TOTEM-arXiv-1712.06153} is obtained for $\sqrt{s} \geq 10$
GeV. For the $pp$ scattering both curves based on the result (\ref{eq:4.b.3}) and on the parameterizations from
\cite{PDG-PRD-98-030001-2018,TOTEM-arXiv-1712.06153} show a gradual
decreasing of the $G_{\scriptsize{\mbox{inel}}}(s,0)$ considering ultra-high
energies $\sqrt{s} \gtrsim 100$ TeV and (\ref{eq:4.b.3}) leading
to the noticeable deviation from black disk limit at
$\mathcal{O}$(100 TeV). In general, this observation is also valid
for $\bar{p}p$ (Fig. \ref{fig:2}). However, in this case the
behavior of the curves is characterized by considerable
uncertainty in multi-TeV energy domain $\sqrt{s} > 10$ TeV due to the
lack of experimental estimations.

\begin{figure}
\centering
\includegraphics[width=17.0cm,height=14.0cm]{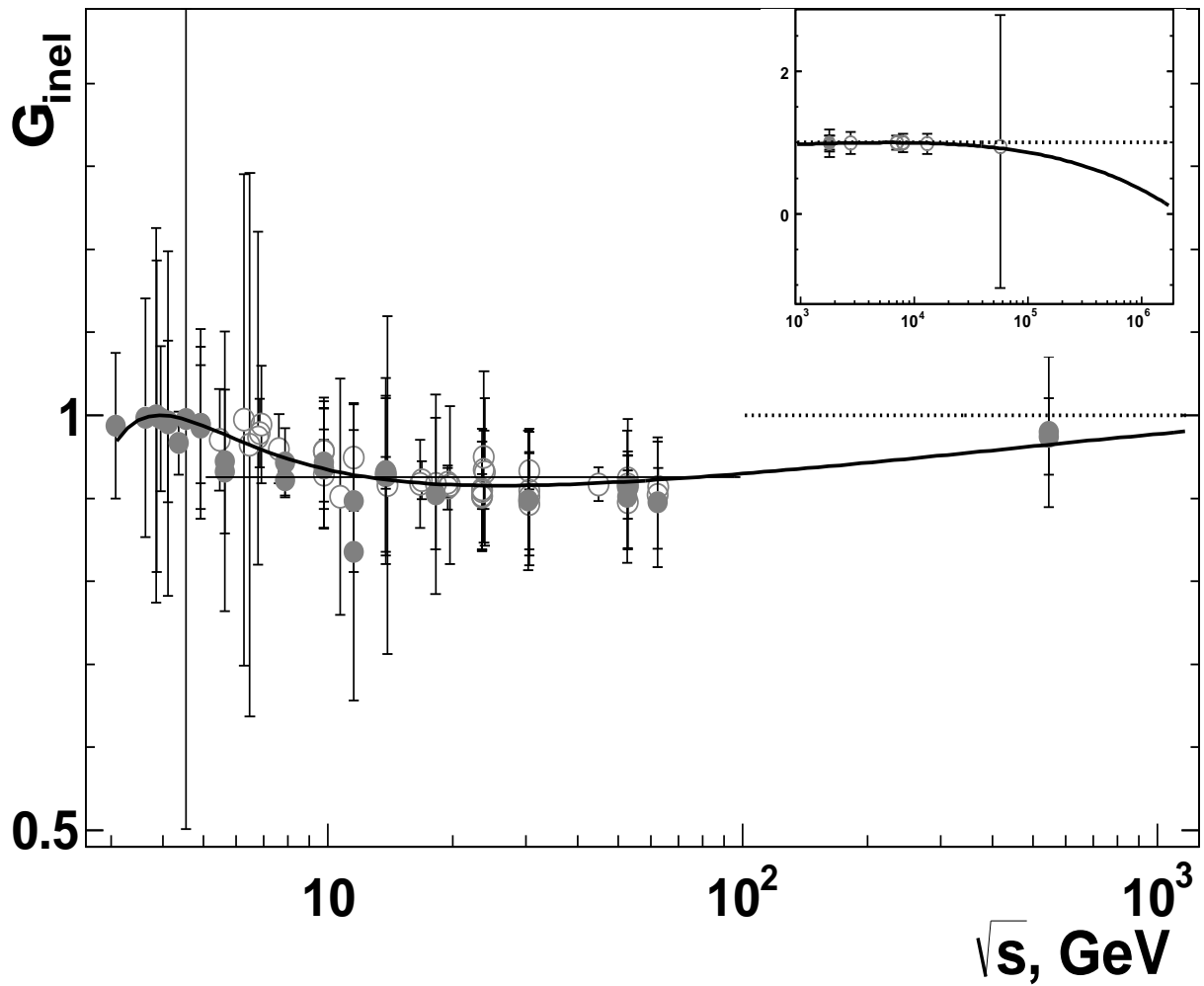}
\caption{Energy dependence for the inelastic overlap function in
central nucleon-nucleon collisions
$G_{\scriptsize{\mbox{inel}}}(s,0)$. Experimental estimations
obtained within present work are shown by open points for $pp$ and
by solid points for $\bar{p}p$ scattering, curves are evaluated
with help of the smooth dependencies for
$R_{\scriptsize{\mbox{e/t}}}$ from
\cite{okorokov-arXiv-1805.10514}. Solid curve corresponds to the
results of the fitting of experimental sample for
$R_{\scriptsize{\mbox{e/t}}}(s)$ joined for $pp$ and $\bar{p}p$ by
(\ref{eq:4.b.3}) at $\sqrt{s_{\scriptsize{\mbox{min}}}}=3$ GeV,
dotted line -- by constant at
$\sqrt{s_{\scriptsize{\mbox{min}}}}=1$ TeV, thin solid curve
corresponds the fit of $R_{\scriptsize{\mbox{e/t}}}$ by constant
in the intermediate energy range at $\sqrt{s} \in [10; 100]$ GeV
(see detailed description in the text). Inner panel: experimental
estimations and curves for the energy domain $\sqrt{s} > 1$ TeV.}
\label{fig:3}
\end{figure}

Fig. \ref{fig:3} shows the $G_{\scriptsize{\mbox{inel}}}(s,0)$ for
nucleon-nucleon collisions. Based on the above discussion, the present results are only shown in Fig. \ref{fig:3} for
clearer picture. The experimental estimations agree for $pp$ and
$\bar{p}p$ scattering for close $\sqrt{s}$. The constant describe
points reasonably for intermediate energies $10 \leq \sqrt{s} \leq
100$ GeV and for high-energy domain. One can note that the
constant dotted line obtained with the help of the corresponding
result for $\sqrt{s} > 1$ TeV from
\cite{okorokov-arXiv-1805.10514} agree quite well with
experimental estimation at smaller $\sqrt{s}=546$ GeV. Then, one
can suggest that the constant allow a reasonable description of
the experimental estimations for
$G_{\scriptsize{\mbox{inel}}}(s,0)$ for joined nucleon-nucleon
sample in wider energy range $\sqrt{s} > 100$ GeV with respect to
the result for $R_{\scriptsize{\mbox{e/t}}}(s)$
\cite{okorokov-arXiv-1805.10514}. The approach based on the
(\ref{eq:4.b.3}) predicts the onset of deviation from the black
disk limit at $\mathcal{O}$(100 TeV) and the continues decreasing of
the inelastic overlap function in central nucleon-nucleon
collisions with the growth of $s$ provides
$G_{\scriptsize{\mbox{inel}}}(s,0) \to 0$ for PeV energies.

At present the Tsallis statistics is mostly used for successful description of the single-particle transverse momentum distribution \cite{EPJA-48-160-2012,PLB-723-351-2013,EPJC-74-2785-2014,AHEP-2015-180491-2015,PRD-91-054025-2015,PRD-92-074009-2015,AHEP-2016-9632126-2016,IJP-90-316-2016,EPJA-53-102-2017}
in various hadron and nucleus collisions in wide energy range. On the other hand, the information is very limited regarding the entropy $S_T$ and its dependence on some kinematic parameters. In this paper, the estimations are obtained for $S_T$ in the impact parameter space and the energy dependence is studied for central $pp$, $\bar{p}p$ collisions.

Taking into account the analysis in \cite{borcsik_mod_phys_lett_a31_1650066_2016} the TE in central $pp$, $\bar{p}p$ collisions is calculated at $\sqrt{s_{c}}=25.0$ GeV in the present work. As seen from (\ref{eq:4.b.2a}), the $\zeta(s)=1.0$ corresponds to a maximum ($s<s_{c}$) or a minimum ($s>s_{c}$) of the $S_{T}(s,0)$. Results from \cite{okorokov-arXiv-1805.10514} show the $\zeta(s) \approx 1.0$ can be reached in separate points at intermediate energies $\sqrt{s} \simeq$ 5 GeV, and this is a characteristic value in TeV energy domain. A detailed analysis of (\ref{eq:4.b.2a}) show that $S_{T}(s,0)$ presents a sharper behavior as $s$ approaches to the critical value $s_{c}$. Furthermore, the absolute values of the TE for $s < s_{c}$ ($|S_{T}|=-S_{T}$) are mostly larger by orders than that for $s \gg s_{c}$ ($|S_{T}|=S_{T}$) at $\sqrt{s_{c}}=25.0$ GeV. Therefore, the $|S_{T}(s,0)|$ seems a more adequate quantity for the study of $s$-dependence in wide energy domain for the symmetric form of the TE in central $pp$, $\bar{p}p$ collisions.

\begin{figure}
\centering
\includegraphics[width=17.0cm,height=14.0cm]{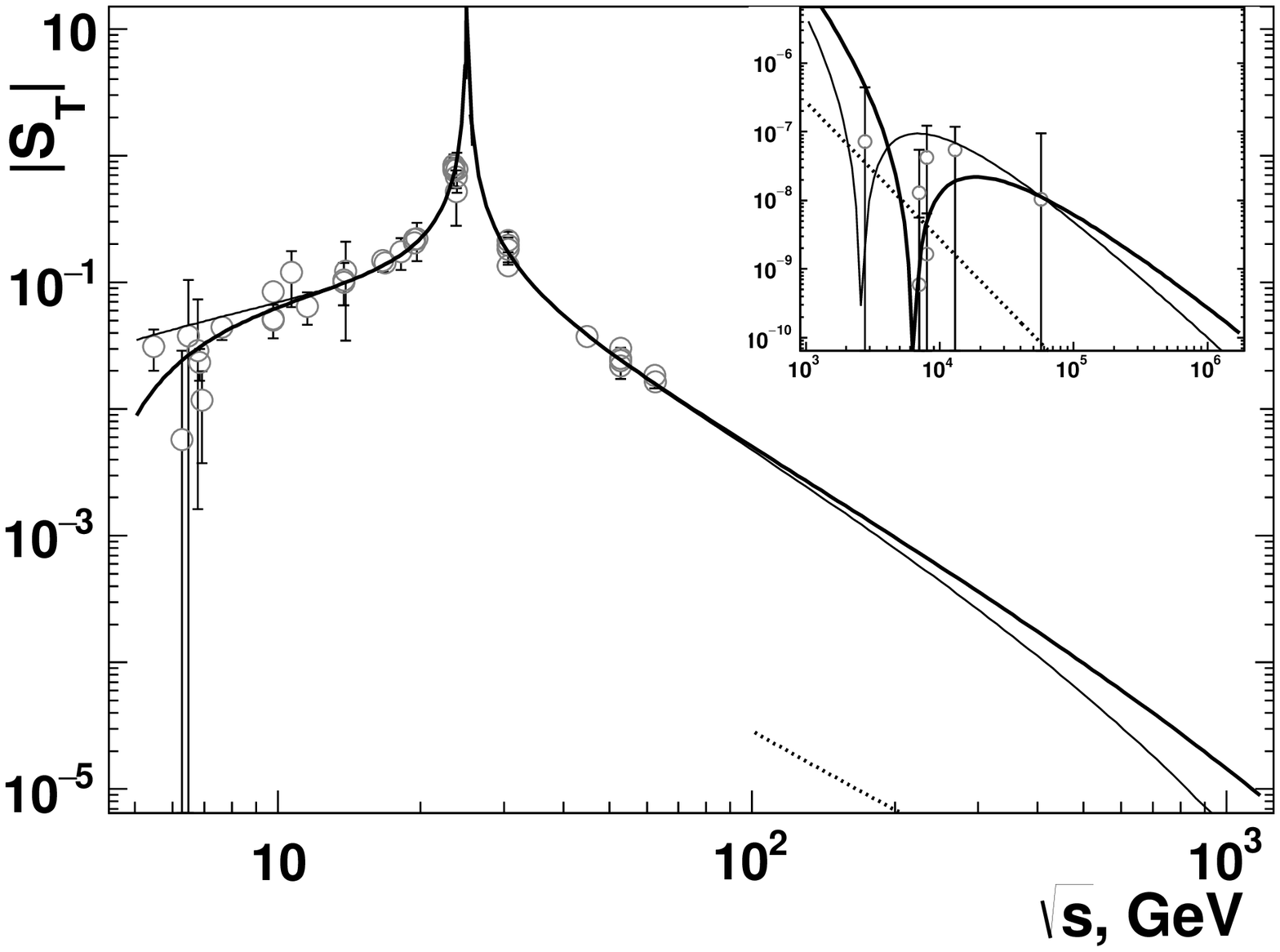}
\caption{Energy dependence for absolute values of the TE in central $pp$ collisions $S_{T}(s,0)$. Notations for experimental estimations and curves are the same as in Fig. \ref{fig:1}. Inner panel: experimental estimations and curves for the energy domain $\sqrt{s} > 1$ TeV.} \label{fig:4}
\end{figure}

\begin{figure}
\centering
\includegraphics[width=17.0cm,height=14.0cm]{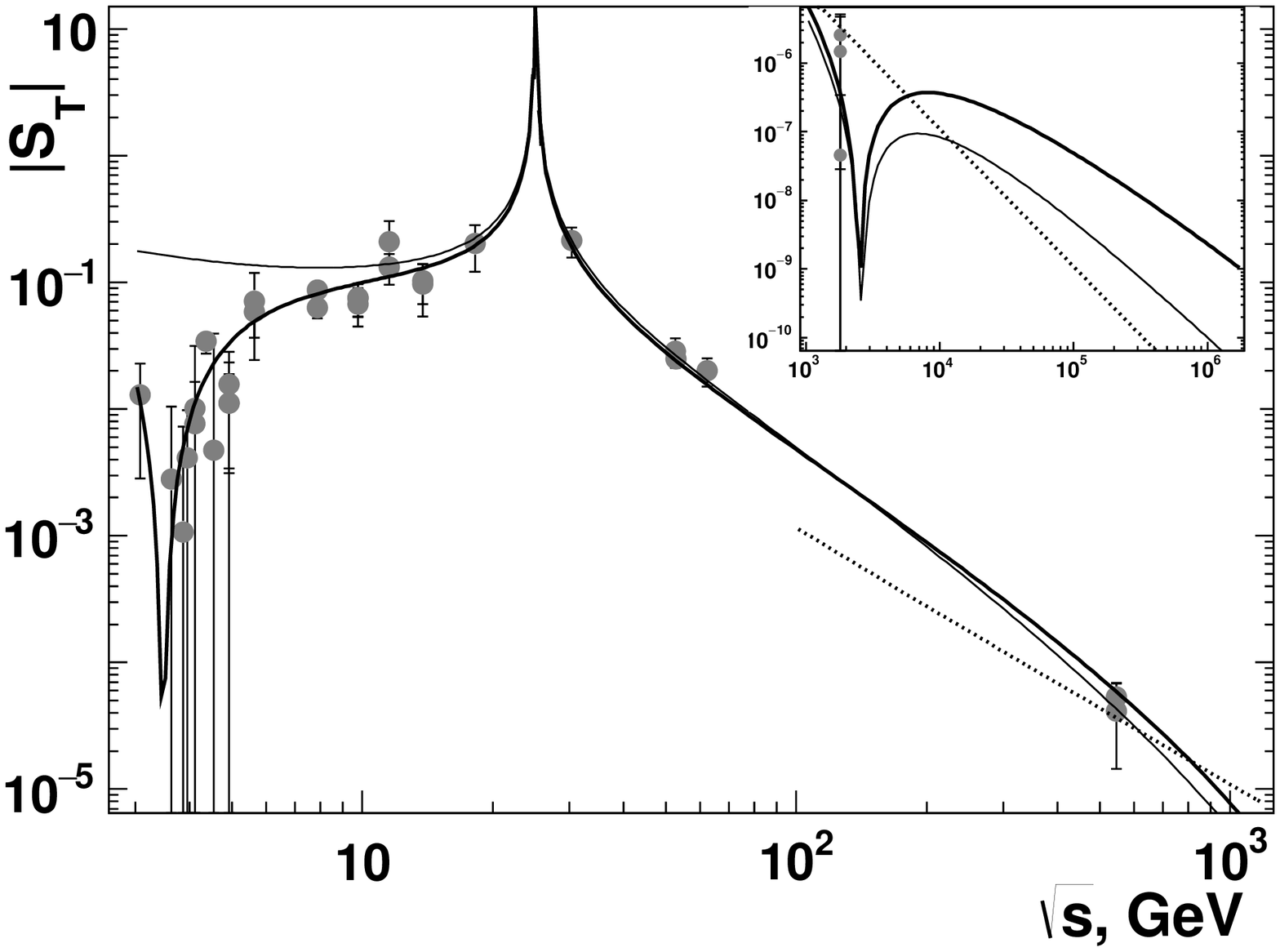}
\caption{Energy dependence for absolute values of the TE in central $pp$ collisions $S_{T}(s,0)$. Notations for experimental estimations and curves are the same as in Fig. \ref{fig:2}. Inner panel: experimental estimations and curves for the energy domain $\sqrt{s} > 1$ TeV.} \label{fig:5}
\end{figure}

\begin{figure}
\centering
\includegraphics[width=17.0cm,height=14.0cm]{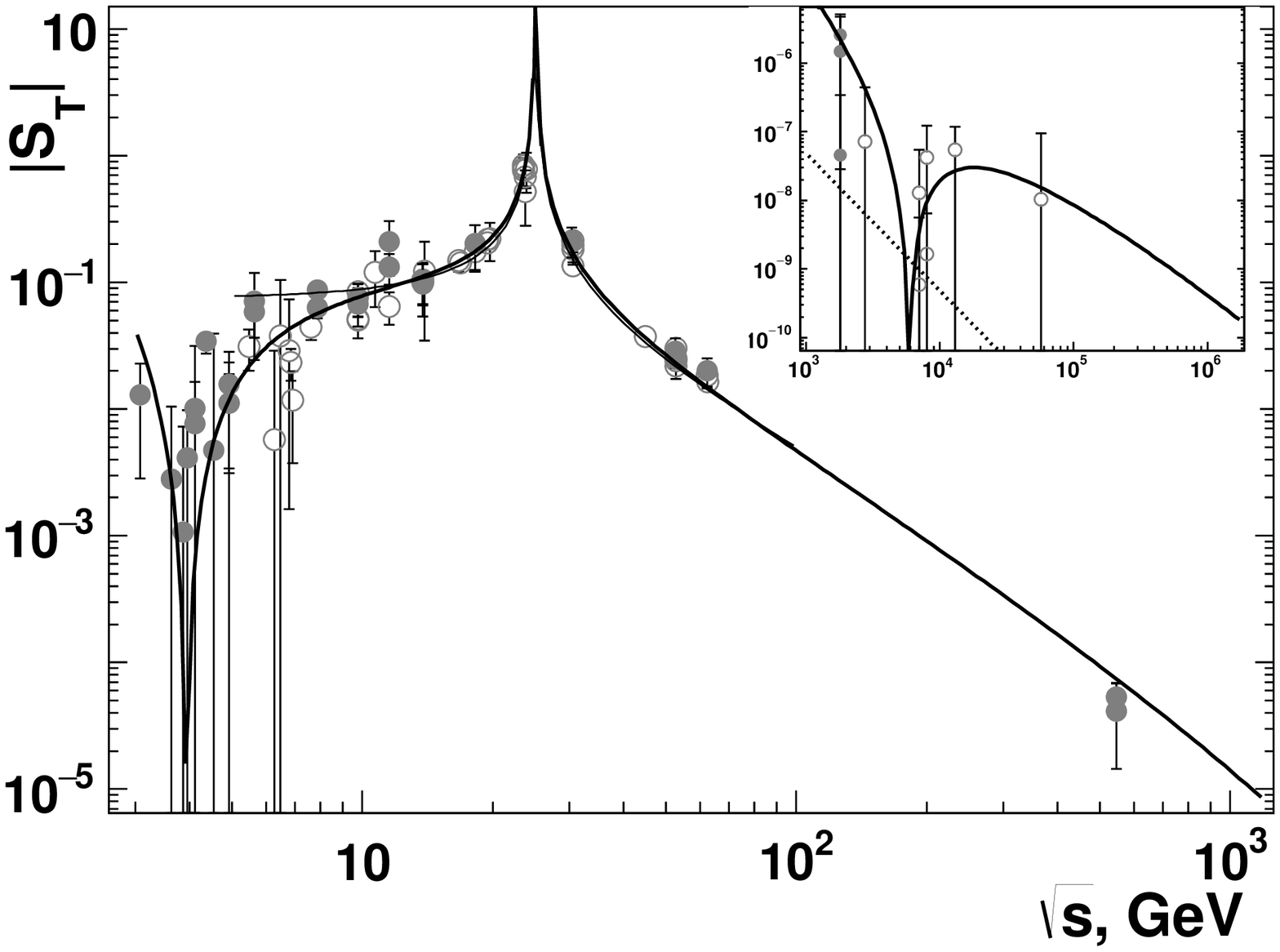}
\caption{Energy dependence for absolute values of the TE in
central $pp$ collisions $S_{T}(s,0)$. Notations for experimental
estimations and curves are the same as in Fig. \ref{fig:3}. Inner
panel: experimental estimations and curves for the energy domain
$\sqrt{s} > 1$ TeV.} \label{fig:6}
\end{figure}

Figs. \ref{fig:4} -- \ref{fig:6} show the energy dependence of the magnitude of the TE in central $pp$, $\bar{p}p$ collisions and for joined sample in nucleon-nucleon scattering, respectively. Experimental estimations for $|S_T(s,0)|$ are deduced with help of the database for $R_{\scriptsize{\mbox{e/t}}}(s)$ from \cite{okorokov-arXiv-1805.10514} and relations (\ref{eq:4.a.3}), (\ref{eq:4.b.2a}). Notations for experimental estimations and smooth curves are the same as in Figs. \ref{fig:1} -- \ref{fig:3}. Maximum value $|S_T(s,0)| \sim 1$ is reached for experimental estimations obtained for $pp$ collisions close to the critical energy $\sqrt{s}_{c}$ (Fig. \ref{fig:4}), different collisions are featured by similar values of $|S_T(s,0)|$ at close values of collision energy (Fig. \ref{fig:6}) and growth of $s$ leads to the fast decreasing of $|S_T(s,0)|$ for $s > s_{c}$. The empirical curves based on (\ref{eq:4.b.3}) and on the parameterizations from \cite{PDG-PRD-98-030001-2018,TOTEM-arXiv-1712.06153} demonstrate the sharp deeps for TeV energies and these deeps are at various $s$ in $pp$ (Fig. \ref{fig:4}) while they coincide in $\bar{p}p$ scattering (Fig. \ref{fig:5}). The parameterizations from \cite{PDG-PRD-98-030001-2018,TOTEM-arXiv-1712.06153} provides $|S_T(s,0)|$ mostly large than the empirical function (\ref{eq:4.b.3}) at low and intermediate energies $s < s_{c}$ for both $pp$ and $\bar{p}p$ collisions. The situation is more ambiguous at high energies $\sqrt{s} > 1$ TeV. For the first case (Fig. \ref{fig:4}), the curve evaluated from (\ref{eq:4.b.3}) lies higher than the curve based on the parameterizations from \cite{PDG-PRD-98-030001-2018,TOTEM-arXiv-1712.06153} up to the $\sqrt{s} \simeq 5$ TeV and in the ultra-high energy domain $\sqrt{s} \gtrsim 100$ TeV. In $\bar{p}p$ scattering (Fig. \ref{fig:5}), the fit result from \cite{okorokov-arXiv-1805.10514} provides smooth curve for $|S_T(s,0)|$, which is higher than the similar curve deduced with the help of the functions from \cite{PDG-PRD-98-030001-2018,TOTEM-arXiv-1712.06153}, and difference is especially visible for $\sqrt{s} \geq 10$ TeV.

%%%%%%%%%%%%%%%%%%%%%%%%%%%%%%%%%%%%%%%%%%%%%%%%%%%%%%%%%%%%%%%%%%%%%%%%
\section{\label{fr}Final Remarks}

As pointed out in \cite{borcsik_mod_phys_lett_a31_1650066_2016}, the total cross section experimental dataset for $pp$ and $\bar{p}p$ present two fractal dimensions. The Peres--Shmerkin theorem \cite{peres_shmerkin} states that if a dataset possesses two fractal dimensions, then the sum of both is not equal to the original dimension of the dataset
\cite{okorokov_int_j_mod_phys_a32_1750175_2017}. The fundamental question is if the total cross section forms a closed, has no isolated points, dense and compact dataset. Of course, the dataset is dense since the general belief is that it can be described by a continuous real-valued function of $s$, for instance. It is compact and has no isolated point as well. However, the term closed implies the existence of a maximum value for the total cross section rise. Note that the Froissart--Martin bound does not prevent this behavior but one cannot assume this from the approach used here.

As noted in \cite{dremin_1,dremin_2,broniowski_arriola} the region near the central collision ($b=0$ fm) presents a growing gray area indicating a tendency for higher energies, corroborated by \cite{fagundes,kfk}. Moreover, the inelastic overlap function is well-described only by the use of at least two Gaussian \cite{fagundes}. This behavior on the impact parameter space may be viewed as a reflex of the occurrence of fractal dimensions in energy and momentum spaces \cite{borcsik_mod_phys_lett_a31_1650066_2016,okorokov_int_j_mod_phys_a32_1750175_2017,bialas_1,bialas_2,antoniou_1,antoniou_2,antoniou_4,antoniou_5}. Therefore, the inelastic overlap function may also present a fractal behavior at each $s$ considered, indicating a phase transition occurring at some $b_0=b(s_c)$.

The principle of maximum entropy states that the probability function \textit{correctly} describing a dataset is the one with the largest entropy $S$. The entropy (\ref{eqnsw_sect4}) is about a particular scattering at some fixed-$s$. To each $s_{i}$ one can construct a dataset taking the pair $\bigl[0,G_{\scriptsize{\mbox{inel}}}(s_{i},b(s_i))\bigr]$, i.e. the line contained in $[0,b(s_i)]$. The dataset thus constructed is a homeomorphism to the Cantor set and then, the Peres--Shmerkin theorem is valid since the dataset formed possess two fractal dimensions. Therefore, by the approach used here the precise knowledge of whole $G_{\scriptsize{\mbox{inel}}}(s,b)$ is avoided by the Peres--Shmerkin theorem and the black disk limit may be reduced to a quasi-black disk limit near $b=0$ (the gray disk in \cite{dremin_1,dremin_2}). This result is independent of the total cross-section reach or not a maximum value.

The TE (\ref{eqnsw_sect4}) can also be related to the amount of information in the area of width $k^{-1}$ and the curve given by $m\bigl[1-nG_{\scriptsize{\mbox{inel}}}(s,b)\bigr]$ depending on each $s$ used. Of course, $m\bigl[1-nG_{\scriptsize{\mbox{inel}}}(s,b)\bigr]$ is limited to the range $\forall\,i: \bigl[0,G_{\scriptsize{\mbox{inel}}}(s_{i},b(s_i))\bigr]$, and, therefore, this area assumes a finite value as well as the amount of information one can obtain from it.

The study of the transition point (the critical temperature) can reveal some important properties of the arrangement of the internal constituents \cite{kapusta_nucl_phys_a148_478_1983}. The temperature at the transition point $s_{c}$ is, of course, of great interest and the result can easily be obtained by using the Helmholtz free energy. The approach considered here entails the possibility of negative temperatures occurring inside the hadron in both energy regimes $s<s_c$ and $s>s_c$. In the first case, the negative temperature allows to hadron the formation of a torus with a smoothed edge toward the center. The latter, indicate the hadron acquires a disk-like shape, tending to a point-like object as the energy tends to infinity. As well-known, the negative temperature has been interpreted as the change in the occupancy of the energy states \cite{nature} along the years: the probability of the occupation of the higher-energy states is greater than the lower-energy states. Therefore, the phase transition obtained here is evidenced by an inversion of the occupation number of the energy states by the internal constituents of the hadron as the energy increases. The negative temperature also avoids the internal constituents to gain kinetic energy, turning the system stable \cite{braun_science_339_52_2013}.

The role of the general entropic index $w=(s/s_{c})^{\alpha}$ in the present model can be enlarged in the Regge theory context. As well-known, in this theory, it is expected that scattering amplitude is dominated by the highest trajectory, $(s/s_c)^\alpha$, where $\alpha$ is momentum-transferred dependent. Of course, the $\alpha$ parameter can be written in $b$-space and, therefore, the Tsallis entropy shows a clear connection with the scattering amplitude in the impact parameter space. Moreover, the cuts in the $J$ plane representing the particle exchange can be studied in terms of the Tsallis entropy simply adopting the general entropic index. Then, the particle exchange contribution to the Tsallis entropy can be taken into account in Regge theory. This study will be performed elsewhere.

%The Hagedorn temperature in its first beginning was thought as the highest possible temperature for the hadronic matter \cite{hagedorn_nuovo_cim_supp_3_147_1965}. However, it was shown later that this temperature is, in fact, a critical phase transition temperature \cite{cabibbo_phys_lett_b59_67_1975} from a hadronic matter to a quark-gluon plasma. The model presented here indicates a phase transition occurring at $b_0=b(s_c)$ for both energy regimes where the hadron suddenly changes its internal configuration in the energy occupation states of its internal constituents. In the asymptotic energy regime, the critical value $b_0$ shrinks to zero. Besides, if the majority of the quarks and gluons are located in the region $b<b_0$, then the energy density grows and the higher-energy occupation states must tend to maximum, may be related with the dynamics flavor \cite{albash_phys_rev_d77_066004_2008}. Moreover, this growth also implies in the rise of the total cross section as $s\rightarrow\infty$.

%The use of the Coulomb or the confinement potential cannot allow one to obtain a precise indication of the critical point. However, the approach adopted here may be used as a first glance to improve the theoretical understanding of $pp$ and $\bar{p}p$ collision processes.

The phenomenological analysis for the inelastic overlap function and for the magnitude of the TE in central collisions allows the following conclusions. The $G_{\scriptsize{\mbox{inel}}}(s,0)$ is close to the black disk limit for $\sqrt{s} \lesssim 5$ GeV and, especially, for TeV energies in both $pp$ and $\bar{p}p$ collisions. There is indication on the $G_{\scriptsize{\mbox{inel}}}(s,0) < 1$ within large errors in the region $10 \lesssim \sqrt{s} \lesssim 100$ GeV. Smooth curves evaluated with the help of the model-independent empirical function (\ref{eq:4.b.3}) and from parameterizations with universal $\ln^{2}\varepsilon$ asymptotic term for
$\sigma_{\scriptsize{\mbox{tot}}}$, $\sigma_{\scriptsize{\mbox{el}}}$
show the deviation of $G_{\scriptsize{\mbox{inel}}}(s,0)$ from the black disk limit for ultra-high energies. The curve based on the model-independent approach predicts $G_{\scriptsize{\mbox{inel}}}(s,0) \to 0$ in nucleon-nucleon collisions for PeV energies. The experimental estimations for TE magnitude reaches the maximum $|S_{T}(s,0)| \sim 1$ close to the critical energy and smooth curves predict very small values of $|S_{T}(s,0)|$ for nucleon-nucleon collisions for ultra-high energies, in particular, $|S_{T}(s,0)| \sim 10^{-10}$ at $\sqrt{s} \sim 1$ PeV in accordance with the model-independent curve based on the equation (\ref{eq:4.b.3}).

%%%%%%%%%%%%%%%%%%%%%%%%%%%%%%%%%%%%%%%%%%%%%%%%%%%%%%%%%%%%%%%%%%%%%%%%
\section*{Acknowledgments}

S.D.C. and C.V.M. thanks to UFSCar by the financial support. The work of V.A.O. was supported partly by NRNU MEPhI Academic Excellence Project (contract No 02.a03.21.0005 on 27.08.2013).

% Figure 2

% Figure 3

% Figure 4

% Figure 5

% Figure 6


\begin{thebibliography}{0}
%1
\bibitem{sachdev_book_2011}S. Sachdev, {\it Quantum Phase Transitions}. Cambridge Univ. Press (2011).
%2
\bibitem{fisher_rep_prog_phys_30_615_1967}M. E. Fisher, Rep. Prog. Phys. {\bf 30}, 615 (1967). %The Theory of Equilibrium Critical Phenomena.
%3
\bibitem{Landau-StatisticalPhysics-1980}
L. D. Landau and E. M. Lifshitz, {\it Statistical physics. Part
1.} Elsevier (1980).
%4
\bibitem{berezinskii_zh_eksp_teor_fiz_59_907_1970} V. L. Berezinskii, Sov. Phys. JETP {\bf 32}, 493 (1971). %Zh. Eksp. Teor. Fiz. {\bf 59}, 907 (1970)
%5
\bibitem{kosterliz_j_phys_c6_1181_1973} J. M. Kosterlitz and D. J. Thouless, J. Phys. C{\bf 6}, 1181 (1973).
%6
\bibitem{tsallis}C. Tsallis, J. Stat. Phys. {\bf 52}, 479 (1988); {\it Introduction to
Nonextensive Statistical Mechanics: Approaching a Complex World}.
Springer (2009).
%7
\bibitem{shannon}C. E. Shannon, Bell S. Tech. J. {\bf 27}, 379 (1948).
%8
\bibitem{renyi}A. R\'enyi, Proceedings of the IV Berkeley Symposium on Mathematics,
Statistics and Probability, p. 547 (1960).
%9
\bibitem{neumann}I. Bengtsson and K. Zyczkowski, {\it Geometry of Quantum States:
An Introduction to Quantum Entanglement}. Cambridge Univ. Press (2006). %(1st ed.).
%10
\bibitem{beck_0902.1235v2}C. Beck, arXiv: 0902.1235v2 [cond-mat.stat-mech] (2009).
%11
\bibitem{nature}L. del Rio et al., Nature {\bf 474}, 61 (2011). %J. {\AA}berg, R. Renner, O. Dahlsten and V. Vedral.
%12
\bibitem{Entropy-19-520-2017}
H. Foroozand and S. V. Weijs, Entropy {\bf 19}, 520 (2017).
%13
\bibitem{jimenez}E. Jim\'enez, N. Recalde, and E. J. Chac\'on, Entropy {\bf 19}, 293 (2017).
%14
\bibitem{zborovsky_int_j_mod_phys_a24_1417_2009}I. Zborovsk\'y and M. V. Tokarev, Int. J. Mod. Phys. A{\bf 24}, 1417 (2009).
%15
\bibitem{vancea}I. V. Vancea, Int. J. Mod. Phys. A{\bf 23}, 4485 (2008).
%16
\bibitem{borcsik_mod_phys_lett_a31_1650066_2016}F. S. Borcsik and S. D. Campos, Mod. Phys. Lett. A{\bf 31}, 1650066 (2016).
%17
\bibitem{okorokov_int_j_mod_phys_a32_1750175_2017} V. A. Okorokov and S. D. Campos, Int. J. Mod. Phys. A{\bf 32}, 1750175 (2017). %; arXiv:1704.02135 [hep-ph] (2017).
%18
\bibitem{A.Bialas.R.Peschanski.Nucl.Phys.B273.703.1986}A. Bialas and R. Peschanski, Nucl. Phys. B{\bf 273}, 703 (1986).
%19
\bibitem{A.Bialas.R.Peschanki.Nucl.Phys.B308.857.1988}A. Bialas and R. Peschanki, Nucl. Phys. B{\bf 308}, 857 (1988).
%20
\bibitem{R.C.Hwa.Phys.Rev.D41.1456.1990}R. C. Hwa, Phys. Rev. D{\b 41}, 1456 (1990).
%21
\bibitem{bialas_1}A. Bialas, Nucl. Phys. A{\bf 545}, 285c (1992).
%22
\bibitem{bialas_2}A. Bialas, Acta Phys. Polon. B{\bf 23}, 561 (1992).
%23
\bibitem{antoniou_1} N. G. Antoniou, F. Diakonos and C. G. Papadopoulos, Phys. Lett. B{\bf 265}, 399 (1991).
%24
\bibitem{antoniou_2}N. G. Antoniou, V. E. Zambetakis, F. K. Diakonos, and N. K. Diakonou, Z. Phys. C{\bf 55}, 631 (1992).
%25
\bibitem{antoniou_4}N. G. Antoniou, F. Diakonos, I. S. Mistakidis, and C. G. Papadopoulos, Phys. Rev. D{\bf 49}, 5789 (1994).
%26
\bibitem{antoniou_5}N. G. Antoniou, N. Davis, and F. K. Diakonos, Phys. Rev. C{\bf 93}, 014908 (2015). %; arXiv:1510.03120 (2015).
%27
\bibitem{deppman_phys_rev_d93_054001_2016}A. Deppman, Phys. Rev. D{\bf 93}, 054001 (2016).
%28
\bibitem{I.Zborovsk.M.V.Tokarev.Phys.Rev.D75.094008.2007}I. Zborovsk and M. V. Tokarev, Phys. Rev. D{\bf 75}, 094008 (2007).
%29
\bibitem{G.Wilk.Z.Wlodarczyk.Phys.Lett.B727.163.2013}G. Wilk and Z. W\l odarczyk, Phys. Lett. B{\bf 727}, 163 (2013).
%30
\bibitem{G.Altarelli.G.Parisi.Nucl.Phys.B126.298.1977}G. Altarelli and G. Parisi, Nucl. Phys. B{\bf 126}, 298 (1977).
%31
\bibitem{A.Deppman.T.Frederico.E.Megias.D.P.Menezes.Entropy.20.633.2018}A. Deppman, T. Frederico, E. Meg\'{i}as, and D. P. Menezes. Entropy {\bf 20}, 633 (2018).
%32
\bibitem{A.Deppman.Adv.High.Ener.Phys.2018.9141249.2018}A. Deppman, Adv. High. Ener. Phys.2018, 9141249 (2018)
%33
\bibitem{RMP-36-655-1964}L. Van Hove, Rev. Mod. Phys. {\bf 36}, 655 (1964).
%34
\bibitem{Collins-book-1977}
P. D. B. Collins, {\it An Introduction to Regge Theory and High Energy Physics}. Cambridge Univ. Press (1977).
%35
\bibitem{Barone-book-2002}
V. Barone and E. Predazzi, {\it High-Energy Particle Diffraction}.
Springer (2002).
%36
\bibitem{sdc2}S. D. Campos, Int. J. Mod. Phys. A{\bf 25}, 1937 (2010).
%37
\bibitem{Tevatron} N. A. Amos et al. (E710 Collaboration), Phys. Lett. B{\bf 247}, 127 (1990); F. Abe et al. (CDF Collaboration), Phys. Rev. D{\bf 50}, 5518 (1994).
%38
\bibitem{totem}G. Antchev et al. (TOTEM Collaboration), Europhys. Lett. {\bf 96}, 21002 (2011).
%39
\bibitem{dremin_1}I. M. Dremin, Phys. Uspekhi {\bf 58}, 61 (2015).
%40
\bibitem{dremin_2}I. M. Dremin, Phys. Uspekhi {\bf 60}, 333 (2017).
%41
\bibitem{broniowski_arriola}W. Broniowski and E. Ruiz Arriola, Acta Phys. Polon. B Proc. Supp., {\bf 10}, 1203 (2017).
%42
\bibitem{alkin_martynov}A. Alkin, E. Martinov, O. Kovalenko, and S. M. Troshin, Phys. Rev. D{\bf 89}, 091501 (2014).
%43
\bibitem{anisovich_nikonov}V. V. Anisovich, V. A. Nikonov, and J. Nyiri, Phys. Rev. D{\bf 90}, 074005 (2014).
%44
\bibitem{troshin_tyurin_1}S. M. Troshin and N. E. Tyurin, Int. J. Mod. Phys. A{\bf 29}, 1450151 (2014).
%45
\bibitem{anisovich}V. V. Anisovich, Phys. Uspekhi {\bf 58}, 1043 (2015).
%46
\bibitem{troshin_tyurin_2}S. N. Troshin and N. E. Tyurin, Mod. Phys. Lett. A{\bf 31}, 1650079 (2016).
%47
\bibitem{albacete_sotoontoso}J. L. Albacete and A. Soto-Ontoso, Phys. Lett. B{\bf 770}, 149 (2017).
%48
\bibitem{arriola_broniowski}E. Ruiz Arriola and W. Broniowski, Phys. Rev. D{\bf 95}, 074030 (2017).
%49
\bibitem{avila_epj_2006} R. F. Avila, S. D. Campos, M. J. Menon, and J. Montanha, Eur. Phys. J. C{\bf 47},
171 (2006).
%50
\bibitem{navarra}F. S. Navarra, O. V. Utyuzh, G. Wilk, and Z. Wlodarczyk, Phys. Rev. D{\bf 67}, 114002 (2003).
%51
\bibitem{PRL-122-120601-2019}P. Jizba and J. Korbel, Phys. Rev. Lett. {\bf 122}, 120601 (2019).
%52
\bibitem{capek_book_2005}V. \v{C}\'apek and D. P. Sheehan {\it Challenges to the Second Law of Thermodynamics: Theory and Experiment}. Springer (2005).
%53
\bibitem{cheng_wu_phys_rev_lett_24_145_1970}H. Cheng and T. T. Wu, Phys. Rev. Lett. {\bf 24}, 145 (1970).
%54
\bibitem{heisenberg_z_phys_133_65_1952}W. Heisenberg, Z. Phys. {\bf 133}, 65 (1952).
%55
\bibitem{PLB-723-351-2013}J. Cleymans et al., Phys. Lett. B{\bf 723}, 351 (2013).
%56
\bibitem{EPJC-74-2785-2014}M. Rybczy$\acute{\mbox{n}}$ski and Z. Wlodarczyk, Eur. Phys. J. C{\bf 74}, 2785 (2014).
%57
\bibitem{AHEP-2016-9632126-2016}H. Zheng and L. Zhu, Adv. High Energy Phys. {\bf 2016}, 9632126 (2016).
%58
\bibitem{EPJA-53-102-2017}A.S. Parvan, O.V. Teryaev and J. Cleymans, Eur. Phys. J. A{\bf 53}, 102 (2017).
%59
\bibitem{rotundo_eur_phys_j_b86_169_2013}G. Rotundo and M. Ausloos, Eur. Phys. J. B{\bf 86}, 169 (2013).
%60
\bibitem{hagedorn_nuovo_cim_supp_3_147_1965}R. Hagedorn, Nuovo Cim. Suppl. {\bf 3}, 147 (1965); Nuovo Cim. A{\bf 56}, 1027 (1968).
%61
\bibitem{dremin}I. M. Dremin, Bull. Lebedev Phys. Inst. {\bf 44}, 94 (2017). %arXiv: 1511.03212 [hep-ph] (2015).
%62
\bibitem{fagundes}D. A. Fagundes, M. J. Menon and P. V. R. G. Silva, Nucl. Phys. A{\bf 946}, 194 (2016). %; arXiv:1509.04108 (2015).
%63
\bibitem{okorokov-arXiv-1805.10514}
V. A. Okorokov, arXiv: 1805.10514 [hep-ph] (2018). %Phys. At. Nucl. {\bf 82}, 134 (2019).
%64
\bibitem{PU-58-963-2015}
V. V. Anisovich, Phys. Uspekhi, {\bf 58}, 963 (2015).
%65
\bibitem{martin_phys_rev_d80_065013_2009}A. Martin, Phys. Rev. D{\bf 80}, 065013 (2009).
%66
\bibitem{PDG-PRD-98-030001-2018}
M. Tanabashi et al. (Particle Data Group), Phys. Rev. D {\bf 98}, 030001 (2018).
%67
\bibitem{PRD-91-076006-2015}
A. Martin and S. M. Roy, Phys. Rev. D{\bf 91}, 076006 (2015).
%68
\bibitem{TOTEM-arXiv-1712.06153}
G. Antchev et al. (TOTEM Collaboration), arXiv: 1712.06153
[hep-ex] (2017).
%69
\bibitem{PRD-14-3092-1976}
D. S. Ayres et al. Phys. Rev. D{\bf 14}, 3092 (1976).
%70
\bibitem{NPB-166-301-1980}
U. Amaldi and K.R. Schubert, Nucl. Phys. B{\bf 166}, 301 (1980).
%71
\bibitem{EPJA-48-160-2012}J. Cleymans and D. Worku, Eur. Phys. J. A{\bf 48}, 160 (2012).
%72
\bibitem{AHEP-2015-180491-2015}H. Zheng and L. Zhu, Adv. High Ener. Phys. {\bf 2015}, 180491 (2015).
%73
\bibitem{PRD-91-054025-2015}L. Marques, J. Cleymans, and A. Deppman, Phys. Rev. D{\bf 91}, 054025 (2015).
%74
\bibitem{PRD-92-074009-2015}H. Zheng, L. Zhu, and A. Bonasera, Phys. Rev. D{\bf 92}, 074009 (2015).
%75
\bibitem{IJP-90-316-2016}Y.--Q. Gao and F.--H. Liu, Indian J. Phys. {\bf 90}, 319 (2016).
%76
\bibitem{peres_shmerkin} Y. Peres and P. Shmerkin, Erg. Theor. Dynam. Syst. {\bf 29}, 201 (2009).
%77
\bibitem{kfk}A. K. Kohara, E. Ferreira, and T. Kodama, Eur. Phys. J. C{\bf 74}, 3175 (2014). %arXiv: 1408.1599 [hep-ph] (2014).
%78
\bibitem{kapusta_nucl_phys_a148_478_1983}J. I. Kapusta and K. A. Olive, Nucl. Phys. A{\bf 408}, 478 (1983).
%79
\bibitem{braun_science_339_52_2013}S. Braun et al., Science {\bf 339}, 52 (2013).

%%%%%%%%%%%%%%%%%%%%%%%%%%%%%%%%%%%%%%%%%%%%%%%%%%%%%%%%%%%%%%%%%%% FIGURES
\clearpage
% Figure 1
\end{thebibliography}
\end{document}